\documentclass[sigconf]{acmart}

\usepackage{tikz}
\usepackage{xspace}
\usepackage{balance}
\usepackage{flushend}
\usepackage{amsmath}
\usepackage{bm}
\usepackage{algorithmic}
\usepackage{graphicx}
\usepackage{caption}
\usepackage{gensymb}
\usepackage{subfigure}
\usepackage{textcomp}
\usepackage{xcolor}
\usepackage{colortbl}
\usepackage{booktabs}
\usepackage{url}

\AtBeginDocument{%
  }

\copyrightyear{2023}
\acmYear{2023}
\setcopyright{rightsretained}
\acmConference[MobiSys '23]{The 21st Annual International Conference on Mobile Systems, Applications and Services}{June 18--22, 2023}{Helsinki, Finland}
\acmBooktitle{The 21st Annual International Conference on Mobile Systems, Applications and Services (MobiSys '23), June 18--22, 2023, Helsinki, Finland}
\acmDOI{10.1145/3581791.3596835}
\acmISBN{979-8-4007-0110-8/23/06}

\newcommand{\SystemName}{Leggiero\xspace}
\newcommand{\revision}[1]{#1}

\begin{document}
\title{\SystemName: Analog WiFi Backscatter with Payload Transparency}

\author{Xin Na}
\email{nx20@mails.tsinghua.edu.cn}
\affiliation{%
  \institution{Tsinghua University}
  \city{Beijing}
  \country{China}
}

\author{Xiuzhen Guo}
\email{guoxiuzhen94@gmail.com}
\affiliation{%
  \institution{Tsinghua University}
  \city{Beijing}
  \country{China}
}

\author{Zihao Yu}
\email{zh-yu17@mails.tsinghua.edu.cn}
\affiliation{%
  \institution{Tsinghua University}
  \city{Beijing}
  \country{China}
}

\author{Jia Zhang}
\email{j-zhang19@mails.tsinghua.edu.cn}
\affiliation{%
  \institution{Tsinghua University}
  \city{Beijing}
  \country{China}
}

\author{Yuan He}
\email{heyuan@tsinghua.edu.cn}
\authornote{Corresponding Author.}
\affiliation{%
  \institution{Tsinghua University}
  \city{Beijing}
  \country{China}
}

\author{Yunhao Liu}
\email{yunhao@tsinghua.edu.cn}
\affiliation{%
  \institution{Tsinghua University}
  \city{Beijing}
  \country{China}
}




\begin{abstract}
Backscatter is an enabling technology for battery-free sensing in today’s Artificial Intelligence of Things (AIOT). Building a backscatter-based sensing system, however, is a daunting task, due to two obstacles: the unaffordable power consumption of the microprocessor and the coexistence with the ambient carrier’s traffic. In order to address the above issues, in this paper, we present \SystemName, the first-of-its-kind analog WiFi backscatter with payload transparency. Leveraging a specially designed circuit with a varactor diode, this design avoids using a microprocessor to interface between the radio and the sensor, and directly converts the analog sensor signal into the phase of RF (radio frequency) signal. By carefully designing the reference circuit on the tag and precisely locating the extra long training field (LTF) section of a WiFi packet, \SystemName embeds the analog phase value into the channel state information (CSI). A commodity WiFi receiver without hardware modification can simultaneously decode the WiFi and the sensor data. We implement \SystemName design and evaluate its performance under varied settings. The results show that the power consumption of the \SystemName tag (excluding the power of the peripheral sensor module) is 30$\mu$W at a sampling rate of 400Hz, which is 4.8× and 4× lower than the state-of-the-art WiFi backscatter schemes. The uplink throughput of \SystemName is sufficient to support a variety of sensing applications, while keeping the WiFi carrier's throughput performance unaffected.
\end{abstract}

\ccsdesc[500]{Hardware~Wireless integrated network sensors}
\ccsdesc[500]{Networks~Network protocol design}

\keywords{Backscatter; Analog; Phase; RF computing}
 
\maketitle

\vspace{-0.2cm}
\section{Introduction}
\begin{figure}
    \centering
    \includegraphics[width=0.45\textwidth]{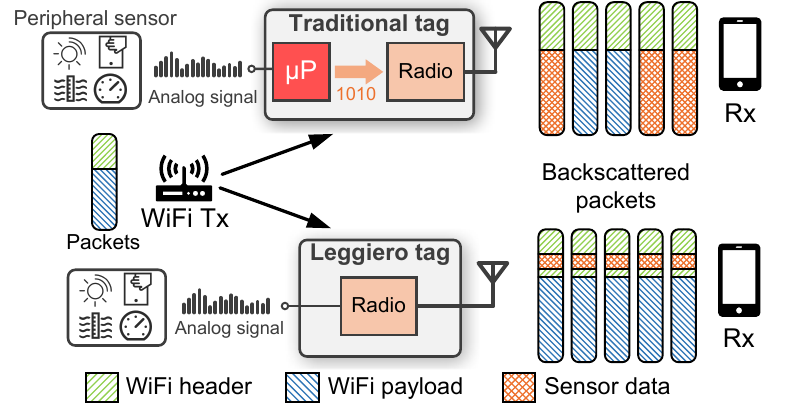}
    \vspace{-0.2cm}
    \caption{\SystemName eliminates the need for using microprocessors ($\bm{\mu}$P) and works transparently with the WiFi carrier's traffic.}
    \label{fig:introduction}
    \vspace{-0.4cm}
\end{figure}
Backscatter is a crucial technology for the Internet of Things (IoT). A backscatter device (i.e., the backscatter tag) is excited by the energy from a carrier source and modulates its own data over the backscattered signals, thus enabling battery-free communication. Sensor data collection is the mainstream application of backscatter. When wired with a sensing module, a backscatter tag becomes a battery-free sensor, delivering the sensor data to the receiver with extremely low power consumption.

Research on backscatter has received broad interest. In recent years, we have witnessed significant progress in this technology with regard to the communication throughput, range, enabled applications, etc. \cite{mmTag, PLoRa, InterTechnologyBackscatter, VideoBackscatter, PassiveZigBee}. But building a backscatter-based sensing system still appears to be a daunting task, mainly due to the following two obstacles:

\textbullet\ \textbf{Unaffordable power consumption of the $\bm{\mu}$P}: Regarding how the sensor data is acquired and transmitted, the conventional practice is to involve a microprocessor ($\mu$P) to interface between the radio and peripheral sensor. Though the power consumption of a backscatter radio can be as low as the level of microwatts ($\mu$Ws), the $\mu$P remains the bottleneck of a sensor's energy consumption. The typical energy consumption of such a $\mu$P is at the level of milliwatts (mWs), which is generally unaffordable, given the stringent energy budget of a battery-free tag \cite{PoweringIoT}.


\textbullet\ \textbf{Coexistence with the ambient carrier's traffic}: The ability to utilize an ambient carrier as the excitation source is critical to the ubiquitous deployment of backscatter, but the other side of the coin is that it is usually hard for the backscatter traffic to coexist with the carrier's traffic. The existing approaches usually need to manipulate the carrier's packets, e.g., by modifying payloads \cite{HitchHike, FreeRider} or corrupting entire frames \cite{WiTAG}, which damages the carrier's traffic and may lead to decoding failure at the receiver.


In order to tackle the above problems, we in this paper propose \SystemName, the first-of-its-kind analog WiFi backscatter scheme. \SystemName directly modulates the analog sensor signals in the CSI (Channel State Information) of the backscattered WiFi packets, which can be received and decoded by a commercial WiFi receiver. Fig. \ref{fig:introduction} compares the schemes of \SystemName and conventional WiFi backscatter. In \SystemName, the sensor is directly interfaced with the radio. The sensor data embedded in the CSI coexist transparently with the payload of the WiFi carrier's packets. The advantages of \SystemName are attributed to the following innovative designs:

\textbullet\ \textbf{Low-power signal conversion in the analog domain.} The analog sensors mostly output signals in the form of voltage. \SystemName takes the sensor's analog voltage as input and directly converts it to the phase of RF signals in the analog domain. We choose the RF phase due to its stability in propagation compared with the amplitude modulation \cite{HybridBackscatter}, as well as its compatibility with the WiFi network compared with the pulse width modulation \cite{VideoBackscatter}. We first show that the reflected RF phase is determined by the reflection coefficient (\S\ref{subsec:reflective_coefficient}). Then we explore the capacitor model (\S\ref{subsec:capacitor_model}) to establish the requirement of the circuit design with respect to the generality, phase range, and conversion linearity. By utilizing the varactor diode, we design a passive reflective circuit (\S\ref{subsec:reflective_circuit_design}) that meets the requirement. Leveraging the varactor diode to alter the reflection coefficient of the tag, \SystemName can control the phase of the reflected signal according to the sensor's analog voltage. In this way, \SystemName avoids using microprocessors as the interfacing media and reduces the energy consumption to an affordable level.

\begin{figure}
    \centering
    \includegraphics[width=0.45\textwidth]{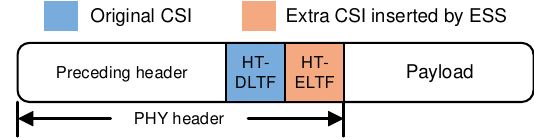}
    \vspace{-0.2cm}
    \caption{Extra spatial sounding featured 802.11n packet. It provides a duplicated CSI since the long-training-fields (LTF) experience the same channel.}
    \label{fig:80211n_packet}
    \vspace{-0.4cm}
\end{figure}

\textbullet\ \textbf{Analog modulation with payload transparency.} \SystemName exploits the ``extra spatial sounding'' (ESS) feature in 802.11n and utilizes CSI to carry the analog sensor signals (\S\ref{subsec:extra_spatial_sounding}). Fig. \ref{fig:80211n_packet} shows the structure of an ESS-enabled packet. The sensor signal, converted into the form of the RF signal phase, is embedded in the extra CSI of a WiFi packet. \SystemName precisely locates the extra long training field (LTF) section of a packet by using an envelope detector, and then embeds the accurate analog phase value by using a carefully designed reference circuit (\S\ref{subsec:embedding_process}). On receiving the backscattered packet, a receiver can efficiently extract the embedded sensor readings as well as cancel out the environmental influences on the CSIs by taking the phase difference of the two CSIs in the same packet (\S\ref{subsec:extracting_messages}). Throughout this process, no modification is made to the payload of the WiFi packet, as is called payload transparency. Hence, the original WiFi data in the payload can also be decoded at the same time. \revision{Importantly, \SystemName does not require hardware modification to commercial WiFi transceivers.}

In addition to the above key design, we also present the MAC layer design of \SystemName (\S\ref{sec:mac_layer}) and the implementation details (\S\ref{sec:implementation}). The hardware schematics are made publicly available.\footnote{Open source hardware can be found at https://github.com/wonderfulnx/Leggiero.} \S\ref{sec:evaluation} presents the comprehensive evaluation of \SystemName. The ASIC power consumption of the \SystemName tag (excluding the power of the peripheral sensor module) is 30$\mu$W at a sampling rate of 400Hz, which is 4.8$\times$ and 4$\times$ lower than the existing WiFI backscatter schemes\cite{FreeRider, WiTAG}. This power enables the sensor tag to work in a battery-free manner. Furthermore, due to the payload transparency of Leggiero, the WiFi carrier’s data traffic is always preserved with unaffected throughput performance. Leggiero achieves 5Kbps throughput, which is sufficient to support a variety of sensing applications. At a high level, \SystemName's design incorporates an RF computing mechanism that operates passively on the RF signal during its propagation.

\S\ref{sec:discussion} discusses practical issues of \SystemName and the potential research space. \S\ref{sec:related_works} briefly introduces related works. We conclude this work in \S\ref{sec:conclusion}.


\section{Analog Signal Conversion}\label{sec:analog_signal_conversion}
This section introduces how \SystemName converts the sensor signal to the phase of the WiFi signal. For a backscatter tag, the phase of the reflected RF signal is determined by the \textbf{reflection coefficient ($\bm{\Gamma}$)} of the tag. Therefore, \SystemName achieves the conversion by relating the sensor signal (usually a voltage signal) with the reflection coefficient. We build a passive RF circuit to produce different $\Gamma$ according to the analog voltage, thus constructing different phases in the reflected signal.
\subsection{Primer: Reflection Coefficient}\label{subsec:reflective_coefficient}
\begin{figure}[t]
    \centering
    \subfigure[\revision{Definition of reflection coefficient $\Gamma$}]{
        \begin{minipage}[t]{0.21\textwidth}
        \includegraphics[width=\textwidth]{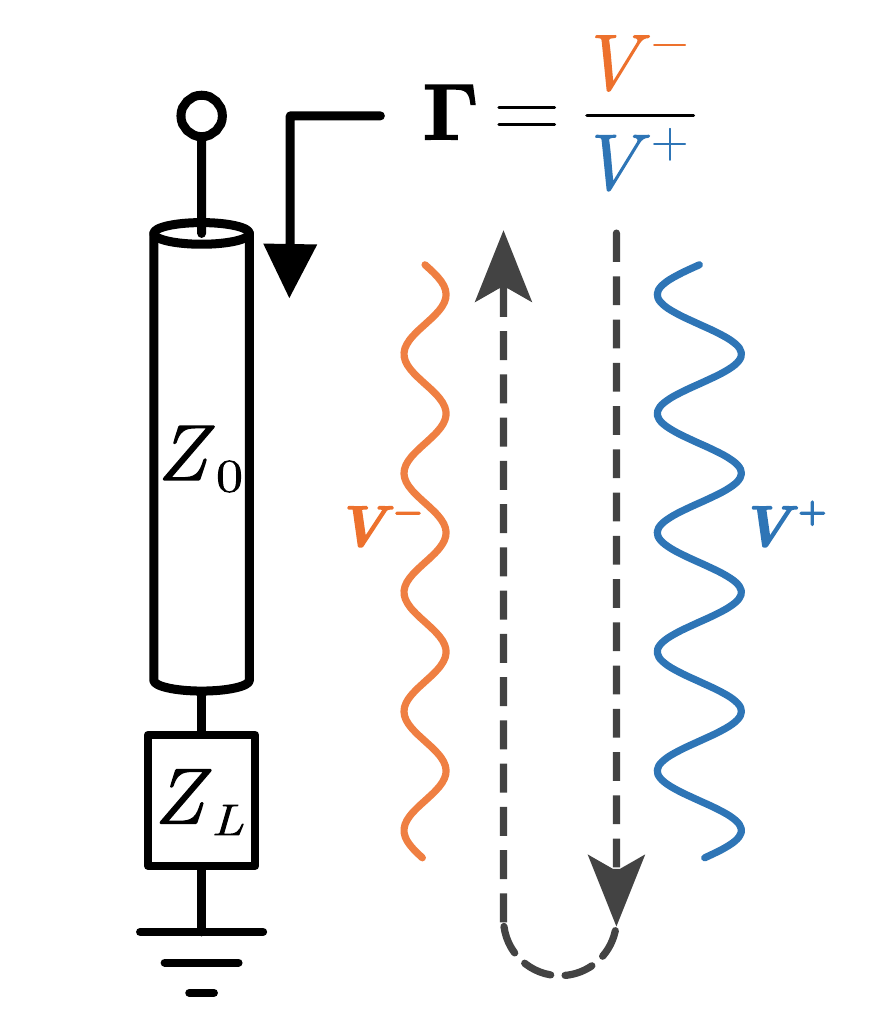}
        \label{fig:reflection_coefficient}
        \end{minipage}
    }
    \hspace{0.2cm}
    \subfigure[$\Gamma$ in smitch chart]{
        \begin{minipage}[t]{0.19\textwidth}
        \includegraphics[width=\textwidth]{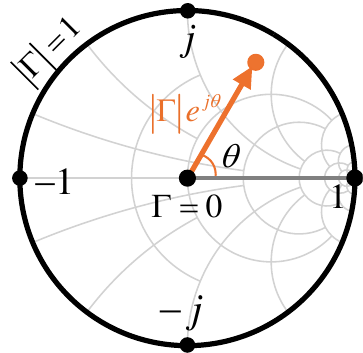}
        \label{fig:smith}
        \end{minipage}
    }
    \vspace{-0.2cm}
    \caption{The reflection coefficient $\Gamma$. (a) shows its definition, where $Z_0$ and $Z_L$ are the impedance of the transmission line and the load, respectively. (b) shows its polar coordinates representation in a Smith chart.}
    \vspace{-0.4cm}
\end{figure}

For an RF circuit, the reflection coefficient represents the ratio of the reflected voltage wave ($V^-$) to the incident voltage wave ($V^+$) at a particular port. Fig. \ref{fig:reflection_coefficient} shows a transmission line of characteristic impedance $Z_0$ feeding a load with impedance $Z_L$. The reflection coefficient $\Gamma$ is given by:
\begin{equation}\label{eq:Gamma}
\Gamma=\frac{V^-}{V^+}=\frac{Z_L-Z_0}{Z_L+Z_0}=|\Gamma|e^{j\theta},
\end{equation}
where $|\Gamma|$ and $\theta$ represent the relative amplitude attenuation and the relative phase variation of the reflected wave compared to the incident wave, respectively. A simple understanding is that for an incident wave signal $A\sin(2\pi ft)$, the reflected signal of this circuit is $A|\Gamma|\sin(2\pi ft+\theta)$. In \textit{S-parameter} theory \cite{pozar2011microwave}, the reflection coefficient is also denoted by $\mathbf{S_{11}}$. In the rest of this paper, we will also use $S_{11}$ to denote the reflection coefficient.

In the conventional backscatter design, the tag modulates bit 0 and bit 1 by providing two discrete $\Gamma$ values. For example, the RFID tag provides $\Gamma_1=0$ as a matched state where the incident wave is completely absorbed and $|\Gamma_2|=1$ as a reflective state where the  wave is completely reflected. These two values can be shown in a Smith chart in Fig. \ref{fig:smith} in polar coordinates. Other examples are recent WiFi backscatter designs such as HitchHike \cite{HitchHike}, the two different values have the same magnitude but provide different phases, e.g., $\Gamma_1=j$ and $\Gamma_2=-j$. It means that the incident signal is always reflected in both states but contains a $180\degree$ phase difference between the two states.

Our insight here is that the reflection coefficient can not only be switched digitally but also be controlled in an analog form, which provides the opportunity to realize signal conversion in the analog domain. By performing an analog variation on the reflection coefficient, the reflected RF phase can be adjusted to carry the tag's sensor readings.

\subsection{Exploring the Capacitor Model}\label{subsec:capacitor_model}
\begin{figure}[t]
    \centering
    \subfigure[Model]{
        \includegraphics[width=0.10\textwidth]{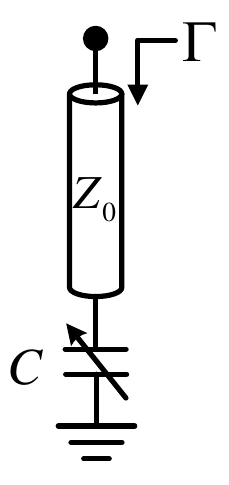}
        \label{fig:circuitmodel}
    }
    \subfigure[Phase range of different capacitance]{
        \includegraphics[width=0.34\textwidth]{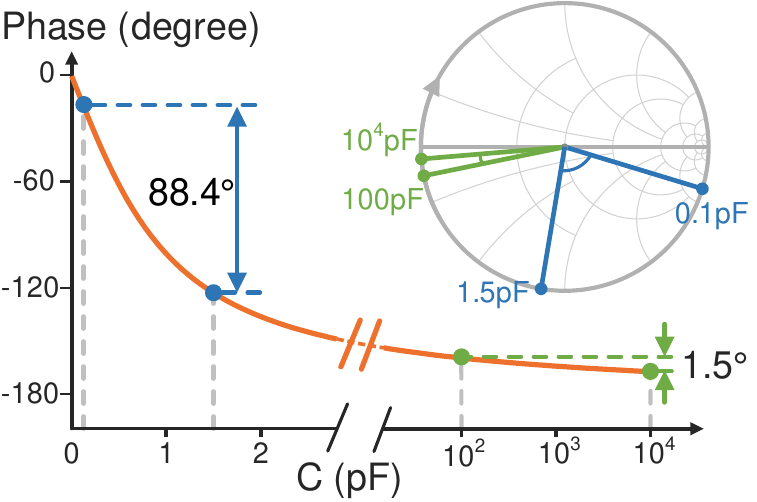}
        \label{fig:rangelinearity}
    }
    \vspace{-0.2cm}
    \caption{The shorted variable capacitor model and the corresponding phase variation. Higher capacitance leads to a smaller phase range and worse linearity.}
    \vspace{-0.4cm}
\end{figure}

A potential method to realize the above-mentioned phase variation is to use a shorted variable capacitor, as shown in Fig. \ref{fig:circuitmodel}, which simply replaces the load of the circuit in Fig. \ref{fig:reflection_coefficient} with a variable capacitor. The reflection coefficient $\Gamma_C$ of such a circuit is computed by replacing the load impedance $Z_L$ with the capacitor impedance $Z_C$ in Eq. \eqref{eq:Gamma}:
\begin{align}\label{eq:capacitorgamma}
    \Gamma_C & = \frac{Z_C-Z_0}{Z_C+Z_0}  = \frac{1-j2\pi f CZ_0}{1+j2\pi f CZ_0}  = e^{j\theta_C}\\
    \label{eq:capacitortheta}
    \theta_C & = -2\arctan(2\pi f C Z_0),
\end{align}
where $f$ is the signal frequency, $C$ represents the capacitance of the capacitor, and $Z_0$ is the characteristic impedance of the antenna or the transmission line, which is usually $50\Omega$. Eq. \eqref{eq:capacitorgamma} shows that the reflection coefficient of a shorted capacitor is a complex unit, and its phase depends only on the capacitance since the signal frequency and the characteristic impedance are constant. Therefore, a shorted capacitor will completely reflect the incident signal, and its capacitance will determine the reflected signal phase.

In order to convert the sensor readings to the RF signal phase, we need to relate the capacitance with the peripheral sensor readings. An intuitive solution is to directly use a capacitive sensor as the shorted variable capacitor. For example, the external pressure on a pressure sensor is directly converted to the variable capacitor's capacitance, which corresponds to an RF signal phase. However, in practice, using a capacitor sensor is not an option due to the following reasons.
\begin{itemize}
    \item \textbf{Generality:} Directly using a capacitive sensor limits \SystemName to only specific sensing scenarios. Such a design cannot support other types of sensor signals since capacitive sensors are a small part of all sensors' designs.
    \item \textbf{Phase Range and Linearity:} The capacitance range of the capacitive sensors is usually above 100pF. According to Eq. \eqref{eq:capacitortheta}, higher capacitance leads to a lower phase range and worse linearity. As shown in Fig. \ref{fig:rangelinearity}, for 2.4GHz WiFi signals, the phase variation between capacitance $0.1$pF and $1.5$pF can be up to $90\degree$, while there is only $1.5\degree$ of the phase difference between $100$pF and $10^4$pF. Note that $\arctan(x)$ function is close to linear when $x\in(0, \frac{\pi}{2})$. Therefore, to achieve a wider phase range and better accuracy, we prefer to control the capacitance between $0$ and $2$pF. Most capacitive sensors cannot satisfy this range requirement.
\end{itemize}
\subsection{Designing Reflective Circuit}\label{subsec:reflective_circuit_design}
Since directly using capacitance as the sensor signal loses generality, we turn to consider that all types of sensor signals can be acquired as voltage signals when sampling. The signal conversion that takes voltage signals as input appears to be a better option. We then need to translate the voltage signal to a variable capacitance. As discussed above, this translation must yield very small capacitance to achieve a wider phase range and better linearity.
\begin{figure*}[htp]
    \centering
    \begin{minipage}[t]{0.32\textwidth}\centering
        \includegraphics[width=0.96\textwidth]{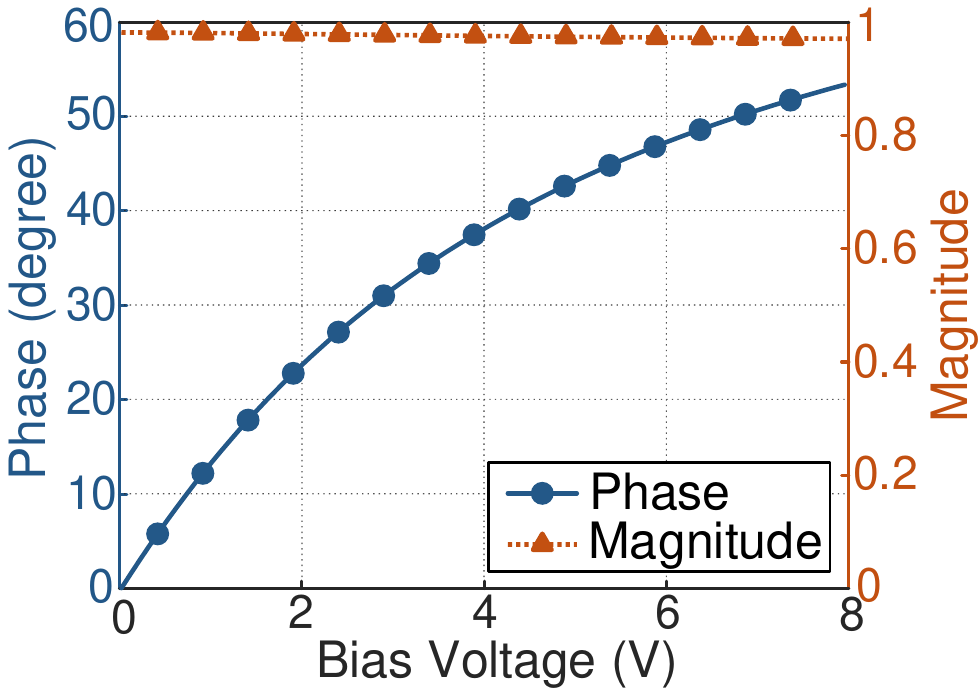}
        \vspace{-0.2cm}
        \caption{Simulated phase and magnitude of the reflected signal v.s. input voltage. The phase is near-linear in 0-5V with little attenuation.}
        \label{fig:phase_vs_vr}
    \end{minipage}
    \hspace{0.2cm}
    \begin{minipage}[t]{0.31\textwidth}\centering
        \includegraphics[width=0.95\textwidth]{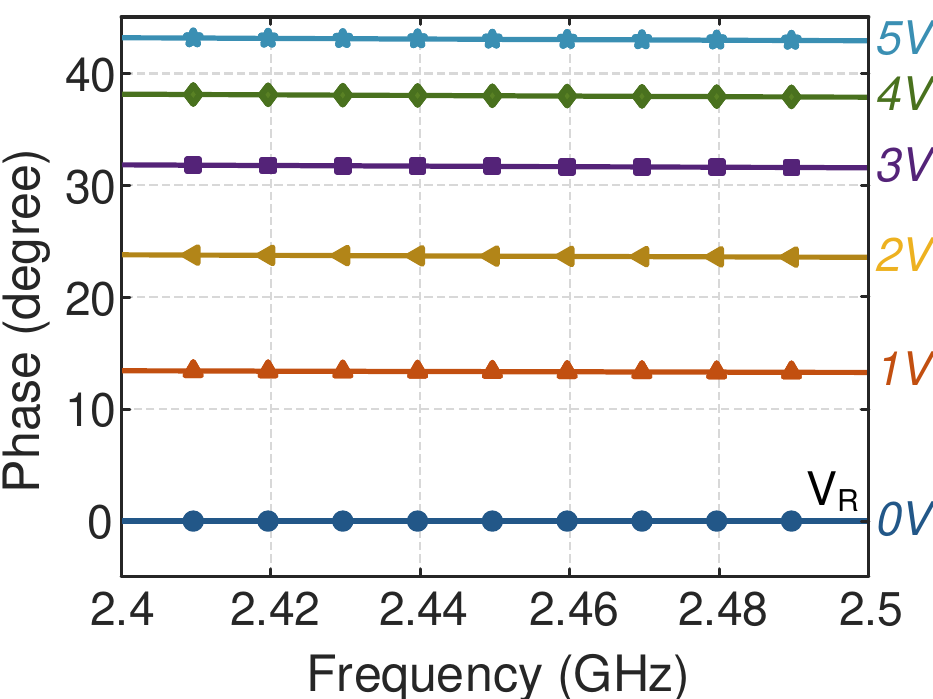}
        \vspace{-0.2cm}
        \caption{Simulated phase v.s. frequency with different input voltage $V_R$. The reflected signal phase is flat on the whole 2.4GHz band.}
        \label{fig:phase_vs_freq}
    \end{minipage}
    \hspace{0.2cm}
    \begin{minipage}[t]{0.32\textwidth}\centering
        \includegraphics[width=0.75\textwidth]{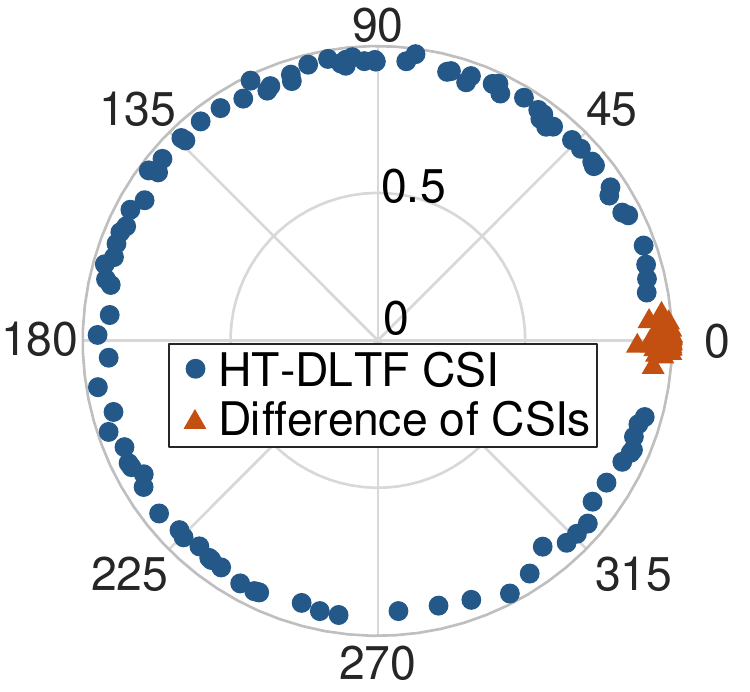}
        \vspace{-0.2cm}
        \caption{Verifying the consistency between the ESS and regular CSI. While their phases vary significantly, the difference remains steady.}
        \label{fig:CSI_consistency}
    \end{minipage}
    \vspace{-0.2cm}
\end{figure*}
\begin{figure}[t]
    \centering
    \subfigure[Capacitance-voltage curve of varactor]{
        \includegraphics[width=0.27\textwidth]{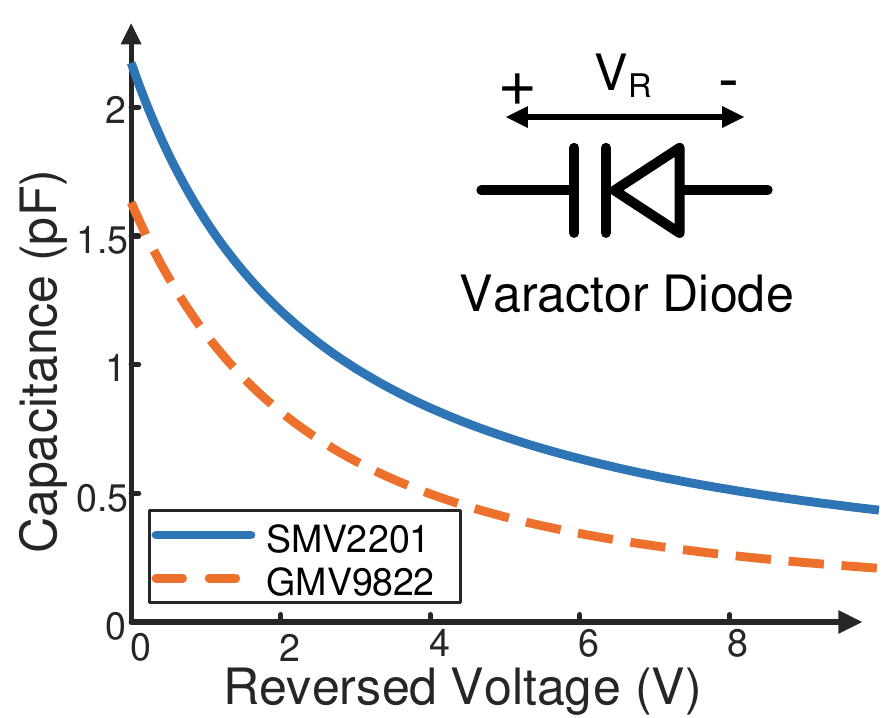}
        \label{fig:varactor}
    }
    \subfigure[Circuit design]{
        \includegraphics[width=0.17\textwidth]{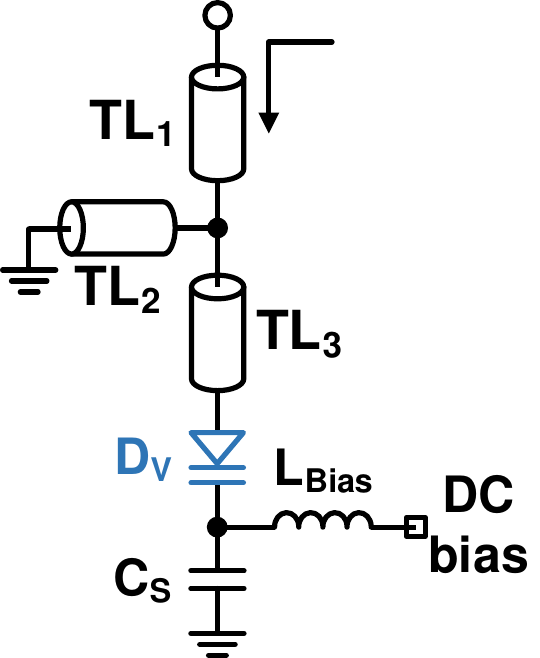}
        \label{fig:circuitdesign}
    }
    \vspace{-0.2cm}
    \caption{\SystemName uses a varactor diode to conduct the analog signal conversion. It converts the input analog voltage to a small capacitance and then shifts the RF phase accordingly.}
    \vspace{-0.4cm}
\end{figure}

\SystemName introduces a varactor diode to convert the external analog voltage into a small capacitance. The varactor diode is a reverse-biased PN junction. It produces a junction capacitance that varies smoothly with the bias voltage. The junction capacitance is dependent on the reversed junction bias voltage, $V$, according to
\begin{align}\label{junction capacitance}
    \vspace{-0.2cm}
    C_j(V)=\frac{C_0}{(1-V/V_0)^\gamma},
    \vspace{-0.2cm}
\end{align}
where $C_0$ is the junction capacitance with no bias; $V_0$ and $\gamma$ depend on the diode type and are constants for a specific diode. Existing commercial varactor diodes can provide the capacitance range of $0$ to $2$pF. For example, a typical GaAs varactor diode can have a junction capacitance that varies from about $0.1$ to $2.0$pF as the reversed bias voltage ranges from $0$ to $20V$.
Specifically, in Fig. \ref{fig:varactor}, we show the junction capacitance versus the voltage for a silicon diode SMV2201 \cite{SMV2201} and a GaAs diode GMV9822 \cite{GMV9822}.

By using the varactor diode, \SystemName relates external analog voltage to the small capacitance and thus the reflection coefficient of the tag. We design a passive circuit as shown in Fig. \ref{fig:circuitdesign}. It provides a continuously variable reflection coefficient that is used to convert the analog voltage signal into the RF phase. The circuit includes three transmission lines TL$_{1,2,3}$, the varactor diode $D_V$, the biasing inductor $L_{Bias}$ that works as a radio frequency choke (RFC) to isolate the DC path and the RF path, and a series capacitor $C_S$ that blocks DC bias and the ground. The shorted to ground transmission line TL$_2$ works as an RF component at 2.4GHz and provides DC for the reverse-biased varactor diode $D_V$. The three transmission lines also offer flexibility in tuning the phase variations, which is evaluated in \S\ref{subsec:tag_component}.

Note that all components used in this circuit are passive components and do not require a power source. The analog signal conversion of \SystemName consumes nearly no power since the main DC path is blocked by the capacitor $C_S$ and the varactor diode $D_V$. The only DC current here is the reverse saturation current of the varactor diode, which is usually less than 0.01$\mu$A. Therefore, \SystemName conducts the analog signal conversion with negligible energy cost.

For a quick proof of the above design methodology, we build and simulate the passive conversion circuit using PathWave Advanced Design System (ADS). By tuning the impedance and the length of the transmission line, the tag provides a flat phase variation over the whole WiFi 2.4GHz frequency band. Fig. \ref{fig:phase_vs_vr} shows the reflected phase versus the bias voltage when the signal frequency is 2.45GHz. We also show the relative phase variation over the whole 2.4GHz band in Fig. \ref{fig:phase_vs_freq}. We can see that the phases are flat and identical over the entire band for a given voltage.

Fig. \ref{fig:phase_vs_vr} also shows the tag's supported input analog voltage range and its resolution (i.e., the amount of phase change corresponding to a certain voltage change). Note that by tuning the length of the three transmission lines TL$_{1,2,3}$ and varying the capacitance of the series capacitor $C_S$, \SystemName tag can trade-off between the voltage range and the resolution. We show the evaluation result of this trade-off in \S\ref{subsec:tag_component}. The resolution of \SystemName also heavily depends on the CSI reception accuracy on the receiver end since the actual sampling part of the voltage takes place in the CSI calculation process. We will assess this accuracy in \S\ref{subsec:analog_conversion_accuracy}.

\revision{\textbf{Novelty of the conversion circuit.}} From the perspective of RF circuit design, \SystemName's reflective conversion circuit is an analog RF phase shifter, except that the circuit is reflective. Existing commercial analog phase shifting RFICs (Radio Frequency Integrated Circuit) can also vary the input RF signal phase according to an analog voltage. \revision{In general, these commercial ICs have complex circuit designs that provide wider frequency and phase-shifting ranges, with some also incorporating varactor diodes for analog-tunable phase shifting. However, we cannot use these components in \SystemName directly; instead, we design our own conversion circuit for the following reasons.}
\revision{\begin{itemize}
    \item \textbf{Power consumption and price:} Commercial ICs are designed for military radars, satellites, and beamforming phase arrays, where performance is the primary concern. For example, HMC928LP5E \cite{HMC928LP5E} is a $450\degree$ analog phase shifter on 2-4GHz. However, the uW-level power consumption demand and pricing constraints of backscatter tags make it impossible to use such ICs, which usually have mW-level power and cost more than \$50 each. Whereas our phase-shifting circuit consumes negligible power and only costs \$2.
    \item \textbf{Tag complexity:} The commercial ICs separate the input and output ports so that the signal enters from one port and exits on another. Using these ICs requires two antennas on the tag, leading to a complicated tag design.
    \item \textbf{Accuracy and robustness:} The robustness of our phase shifting circuit are on par with commercial ICs, as validated in the evaluation with a vector network analyzer (VNA). The result is depicted in the blue line in Fig. \ref{exp:conversion_comparison}, indicating accurate and stable phase shifts conducted on the tag, with no errors.
\end{itemize}}
\revision{In summary, our reflective phase shifter is a simple yet tailored solution that meets all the requirements of backscatter tags while providing accurate phase shifting at the same time.}


\section{Transparent Phase Embedding}\label{sec:selective_phase_embedding}
\begin{figure}
    \centering
    \includegraphics[width=0.45\textwidth]{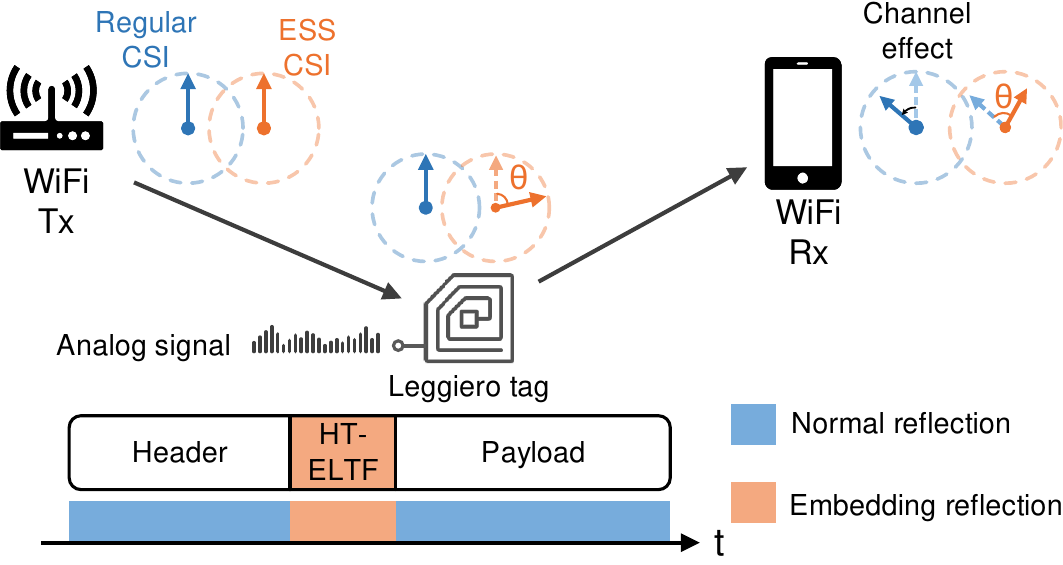}
    \vspace{-0.2cm}
    \caption{Transparent phase embedding of \SystemName. It finds the Extra LTF section of 802.11n packets and embeds the analog sensor reading in the RF phases. It is then extracted by calculating the phase difference.}
    \label{fig:csi_modulation}
    \vspace{-0.4cm}
\end{figure}
We now introduce how \SystemName transparently embeds the converted RF phase information into the WiFi's packet, so that a commercial WiFi device can decode the analog sensor readings and the WiFi data simultaneously. \SystemName exploits the ESS feature of the 802.11n standard and embeds the sensor readings in WiFi's CSI. This section presents the details of the process of phase embedding.
\subsection{Primer: Extra Spatial Sounding}\label{subsec:extra_spatial_sounding}
For the WiFi transmission, CSI \cite{CSISurvey} acts as an indicator of the wireless channel from the transmitter to the receiver at certain frequencies. It characterizes the wireless channel and helps the WiFi receiver to decode the packets. In the backscatter scenario, the tag also influences the wireless channel by introducing attenuation and phase change to the RF signal. \SystemName leverages such influences and modifies the phase of the CSI according to the analog sensor readings.

As we all know, the environment dynamics (including multipath propagation) affect the state of a wireless channel \cite{PicoScenesPaper}. The CSI changes caused by environment dynamics are likely to overwhelm the intentionally embedded phase change of CSI in \SystemName. In order to obtain the phase change correctly, it is necessary to avoid the environmental influences completely.

\SystemName exploits the extra spatial sounding (ESS) feature in 802.11n standard \cite{80211nStd} to cancel out the environmental influences. ESS is originally used to sound extra spatial dimensions (i.e., extra channels) of the multi-input multi-output (MIMO) channel that are not utilized to transmit WiFi data. It inserts the same long training field (LTF) to the physical layer header of a WiFi packet. As shown in Fig. \ref{fig:80211n_packet}, the ESS LTF (or \textbf{HT-ELTF}, \textbf{E} for \textbf{E}xtra) follows closely after the regular HT-DLTF (\textbf{D} for \textbf{D}ata) in the 802.11n preamble, containing the same baseband signal. In a single-input single-output (SISO) scenario, these two LTFs will experience the same channel, thereby giving two identical CSI measurements. To show this consistency, we measure the CSIs in a noisy environment using two QCA9300 WiFi NICs \cite{9300NIC}, as shown in Fig. \ref{fig:CSI_consistency}. As the phase of each CSI changes rapidly, the phase difference is always very close to 0. 
\subsection{Embedding Process}\label{subsec:embedding_process}
In order to embed the sensor reading into a WiFi packet, the \SystemName tag precisely embeds the converted RF phase information in the HT-ELTF section of an ESS-enabled WiFi packet. Other sections of the WiFi packet act as a reference and are reflected with a constant phase, including the original HT-DLTF. We refer to the tag's state when reflecting the HT-ELTF section as \textbf{the embedding state} and the corresponding extra CSI measurement as \textbf{the ESS CSI}. Similarly, we refer to the tag's state when reflecting other sections as \textbf{the reference state} and the CSI as \textbf{the regular CSI}. In this way, the phase difference between the ESS CSI and the regular CSI should be equal to the converted RF phase variation we introduced in \S\ref{sec:analog_signal_conversion}. The environmental influences are completely canceled out when calculating the difference because both the ESS CSI and the regular CSI experience an identical wireless channel. The embedding process is shown in Fig. \ref{fig:csi_modulation}. To realize this design in practice, there are three critical problems to address.

\textbullet\ \textbf{Avoiding self interference.} \SystemName embeds the converted RF phase on the WiFi's CSI by changing the phase of the wireless channel. If the original link from the transmitter to the receiver persists, there exists some signal path that does not include the backscatter tag and results in a confused CSI phase difference. We solve this problem by shifting the frequency of the reflected signal by $\pm$20MHz. The receiver will receive the WiFi packet in a secondary WiFi channel. \revision{This frequency shifting is achieved by multiplying the incident signal with a 20MHz square wave which can be generated using a ring oscillator. Existing WiFi backscatter works, such as HitchHike \cite{HitchHike} and FreeRider \cite{FreeRider}, have used similar approaches. However, they use frequency shifting to separate the backscattered traffic from the carrier traffic's channel. Consequently, a single receiver can only receive either the carrier or the backscattered traffic at a time. To receive both simultaneously, two receivers are necessary. In contrast, \SystemName embeds the sensor signal in the extra CSI section without modifying the payload. In this way, after frequency shifting, the receiver can receive both the backscattered traffic and the carrier traffic simultaneously in the secondary channel without self-interference, improving channel utilization.}

\begin{figure}
    \centering
    \includegraphics[width=0.45\textwidth]{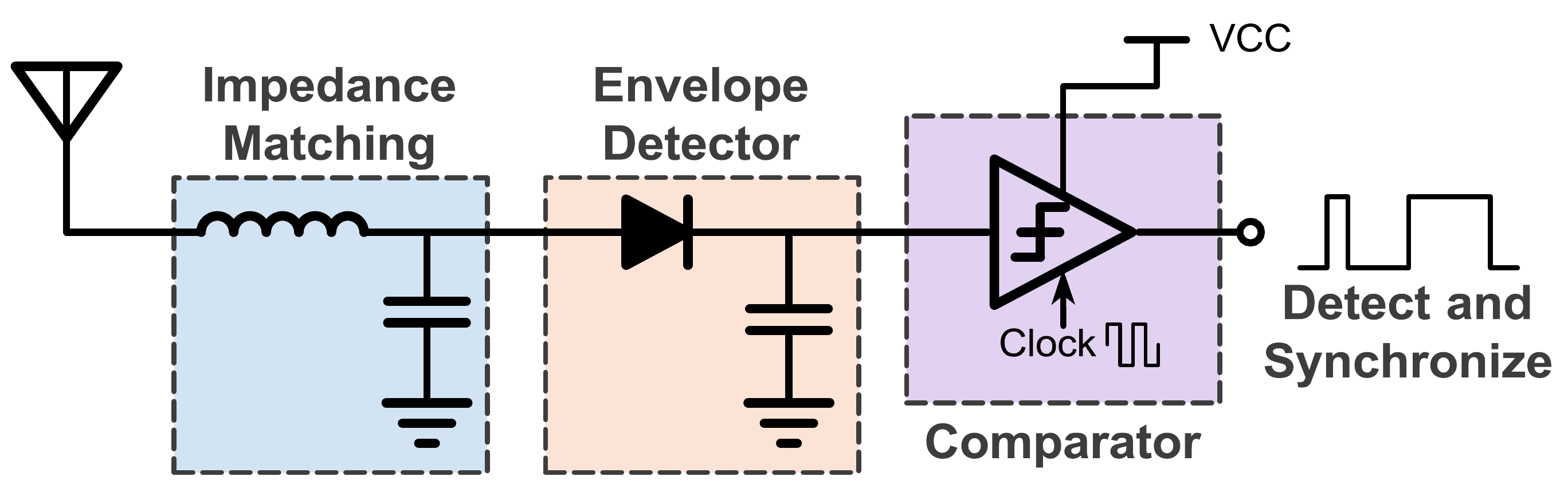}
    \caption{\revision{Packet detection and synchronization circuit of \SystemName. An envelope detector is connected to the antenna. The output signal strength is compared with a threshold to locate the beginning of a packet.}}
    \label{fig:sync_circuit}
    \vspace{-0.4cm}
\end{figure}

\textbullet\ \textbf{Locating HT-ELTF.} Since the \SystemName tag goes into the embedding state only in the HT-ELTF section of a WiFi packet, it needs to synchronize the switching time with this section. \revision{We achieve this by incorporating a commonly used packet detection circuit into the tag, as illustrated in Fig. \ref{fig:sync_circuit}. The circuit consists of an envelope detector in a cascade connection with a voltage comparator. Specifically, the signal strength output of the envelope detector is directly compared to a reference voltage using the comparator. Once the WiFi transmitter starts its transmission, a signal will be generated on the comparator, enabling the tag to identify the beginning of a WiFi packet. Meanwhile, the tag stays in the reference state in the preceding header, which lasts for exactly 36$\mu$s. It then switches to the embedding state in the HT-ELTF section, which has the same 4$\mu$s duration as the WiFi data symbols. Finally, at the end of the HT-ELTF section, the tag switches back to the reference state. Note that the envelope detector can be passive and consumes zero power, allowing the power consumption of this circuit to be as low as a single-digit $\mu$W.}

\revision{In practice, mismatches in the synchronization exist and may impact the performance of the phase embedding. For instance, the tag may experience a delay in its switching time compared to the beginning of the HT-ELTF section. In \SystemName, we find that a synchronization accuracy of 250ns between them is enough for the phase embedding process and results in negligible demodulation errors. This mismatch or tag delay tolerance is mainly attributed to the preceding guard interval (GI) inside the CSI section. The GI lasts for 0.8$\mu$s and is excluded during the CSI calculation, while the latter 3.2$\mu$s baseband signal is actually used. Small synchronization delays (less than 0.8$\mu$s) will fall into the preceding GI section so that its impact is naturally minimized during the GI removal. The 250ns synchronization requirement is met by applying a 4MHz clock to the comparator, which can be derived from the main 20MHz clock. \S\ref{subsec:synchronization_error} presents our validation of the synchronization error (i.e., mismatch) and its impact on backscattering. It shows that the mismatch's impact on demodulation is negligible with the 4MHz clock applied in our implementation. A Similar level of accuracy has also been achieved in state-of-the-art WiFi backscatter systems \cite{HitchHike, FreeRider, SyncScatter}.}

\textbullet\ \textbf{Designing reference circuit.} \SystemName uses the CSI phase difference between the embedding and the reference states to encode and decode the analog sensor readings. The phase in the reference state is also determined by the tag. Therefore, designing the reference circuit is important. Naturally, we set the tag's phase in the reference state equal to a 0V phase in the embedding state. Then, a phase difference of $0\degree$ corresponds to 0V of the tag's analog voltage. The other phase-voltage correspondences are the same as in Fig. \ref{fig:phase_vs_vr}.
\begin{figure}
    \centering
    \includegraphics[width=0.45\textwidth]{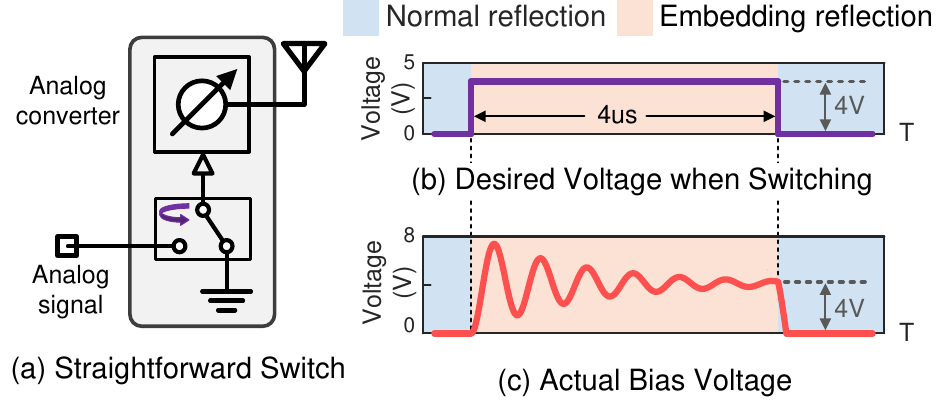}
    \vspace{-0.2cm}
    \caption{Problem of the straightforward approach. (a) The direct voltage switching method. (b) Due to the transient process of the LC circuit, the varactor's actual bias voltage varies in the phase embedding state.}
    \label{fig:transient_process}
\end{figure}
\begin{figure}
    \centering
    \includegraphics[width=0.45\textwidth]{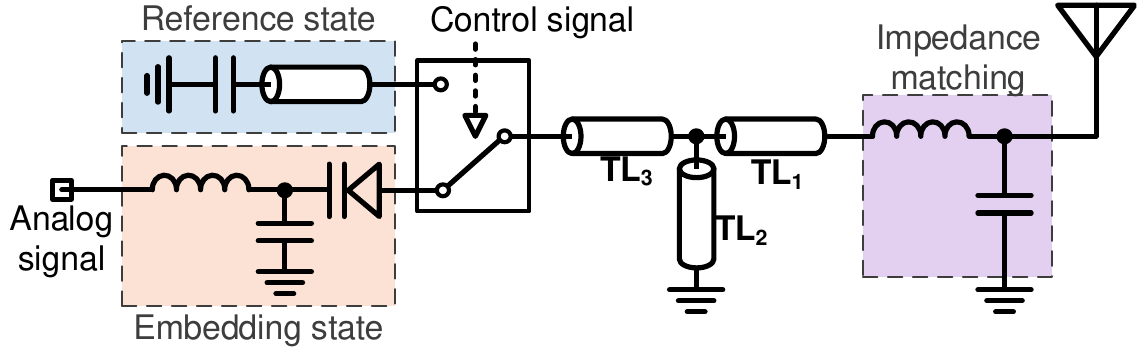}
    \vspace{-0.2cm}
    \caption{Phase embedding circuit schematic of \SystemName. The tag uses an RF switch for the reference circuit.}
    \label{fig:rf_switch}
    \vspace{-0.4cm}
\end{figure}

A straightforward reference circuit design is to switch the DC bias voltage of the varactor diode between the input voltage in the embedding state and 0V in the reference state, respectively, as shown in Fig. \ref{fig:transient_process}(a). However, it does not work in practice. The RF choke $L_{Bias}$ and capacitor $C_S$ together form an LC circuit for the input voltage. This circuit has a transient process when switching from 0V to the input voltage. It causes the voltage of the varactor diode to stabilize gradually instead of changing instantaneously. This transient process may last for more than 2$us$ long in our implementation. It seriously affects the synchronization accuracy and corrupts the embedded phase, as shown in Fig. \ref{fig:transient_process}(b).
\begin{figure*}[htp]
    \centering
    \begin{minipage}[t]{0.28\textwidth}\centering
    \includegraphics[width=0.94\textwidth]{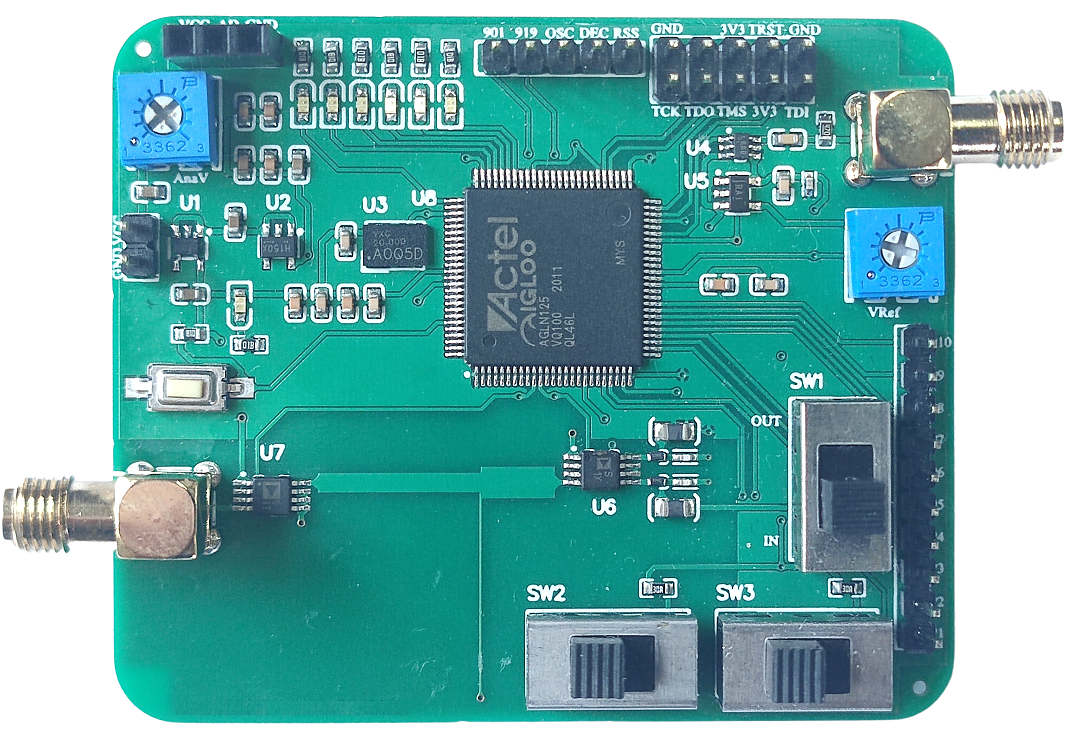}
    \caption{The prototype of \SystemName tag.}\label{fig:tag_implementation}
    \end{minipage}
    \hspace{0.5cm}
    \begin{minipage}[t]{0.23\textwidth}\centering
    \includegraphics[width=0.94\textwidth]{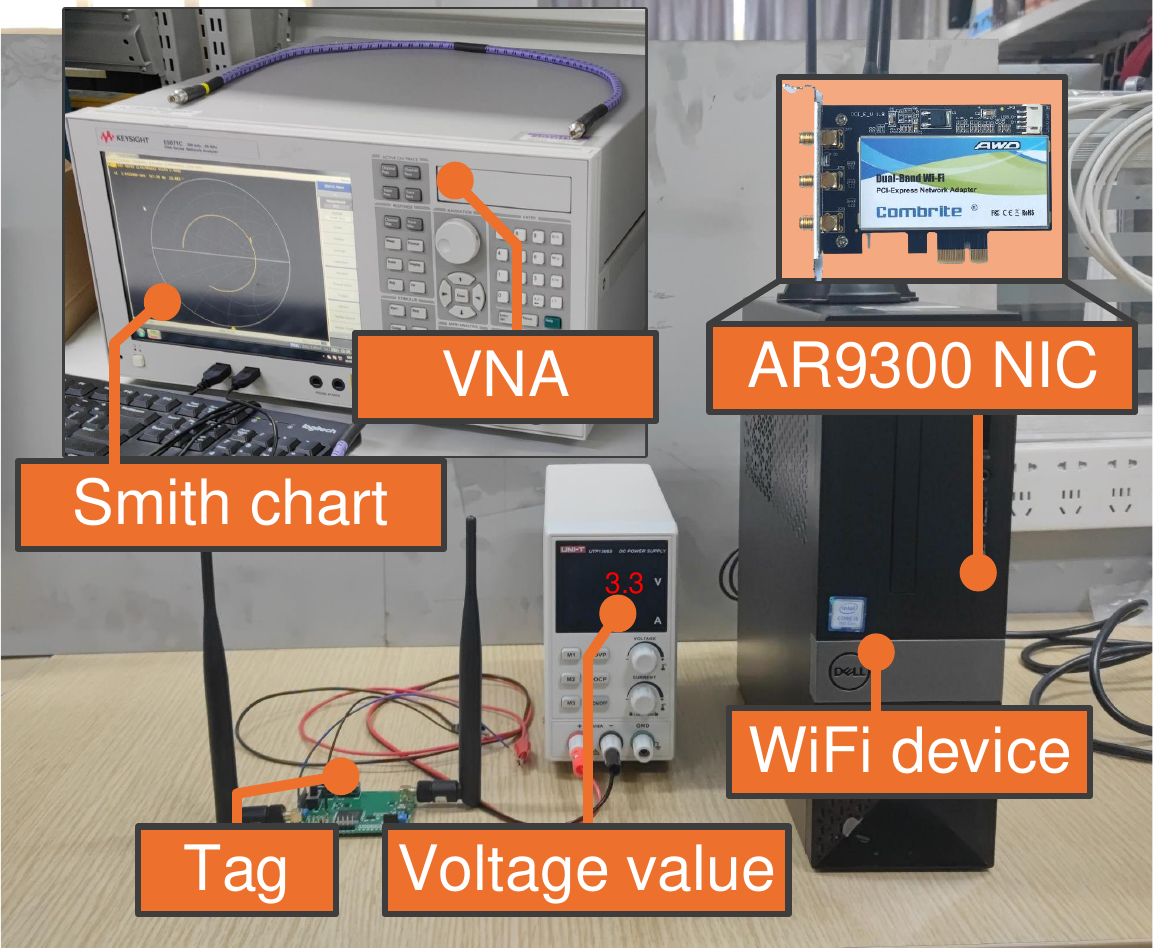}
    \caption{Experiment setup.}\label{fig:experiment_setup
    }\end{minipage}
    \hspace{0.5cm}
    \begin{minipage}[t]{0.40\textwidth}\centering
    \includegraphics[width=0.96\textwidth]{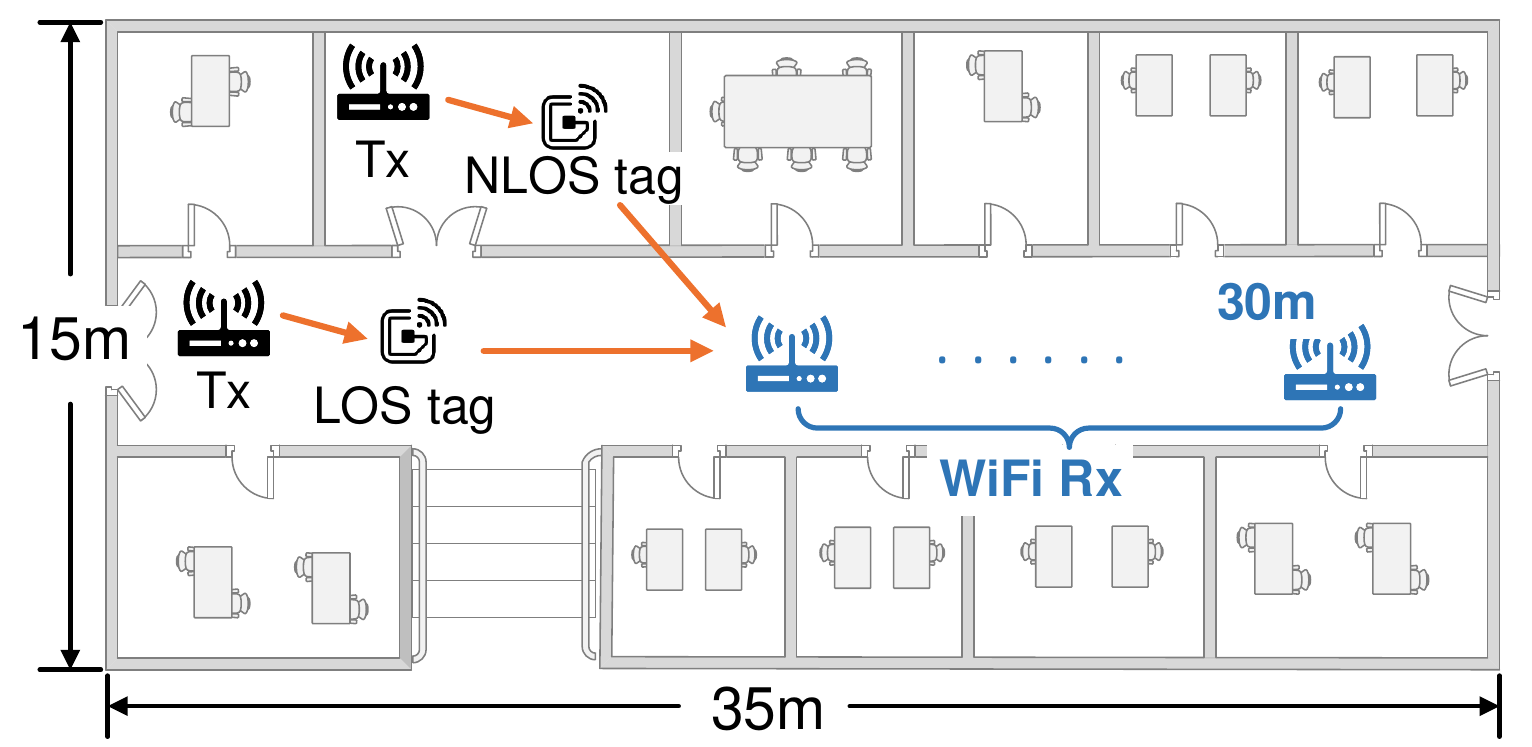}
    \caption{Experiment environment for LOS and NLOS evaluation}\label{exp:floor_plan}
    \end{minipage}
    \vspace{-0.2cm}
\end{figure*}

In order to avoid this transient process and reduce the switching time, \SystemName adds an RF switch in front of the varactor diode. As shown in Fig. \ref{fig:rf_switch}, The RF switch toggles between the varactor diode branch and the constant phase branch, which correspond to the embedding state and the reference state, respectively. By carefully selecting the capacitor and length of the transmission line on the constant phase branch, the \SystemName tag sets the reference state's phase to be equal to a 0V phase of the embedding state. Compared with the straightforward approach, a common RF switch such as ADG918/919 \cite{ADG919} has a switching time of less than 10$ns$, which is more than 200 times faster than the LC transient process. Therefore, using an RF switch to toggle from two branches can easily satisfy our synchronization requirement.
\subsection{Extracting Analog Readings}\label{subsec:extracting_messages}
So far, we have explained how \SystemName embeds the analog sensor readings into a WiFi packet. We now introduce the method to extract these readings.

In the backscatter process of \SystemName, the ESS CSI and the regular CSI experience identical wireless channels except for the phase variation brought by the embedding state of the tag. We denote the phase difference between the embedding state and the reference state as $\theta_V$. Then, the regular CSI $\mathbf{H}_{regular}$ and the ESS CSI $\mathbf{H}_{ess}$ are calculated by:
\begin{align}
    \mathbf{H}_{regular}&=\mathbf{H}_{air}\cdot\mathbf{H}_{err},\\
    \mathbf{H}_{ess}&=\mathbf{H}_{air}e^{j\theta_V}\cdot\mathbf{H}_{err},
\end{align}
where $\mathbf{H}_{air}$ denotes the environmental wireless channel response, and $\mathbf{H}_{err}$ includes all phase errors induced by carrier frequency offset (CFO), sampling frequency offset (SFO), etc. By taking the division of these complex CSI values, we obtain the phase difference converted from the analog voltage:
\begin{equation}
    \frac{\mathbf{H}_{ess}}{\mathbf{H}_{regular}}=e^{j\theta_V}\,.
\end{equation}

After obtaining the phase difference $\theta_V$, we should convert it back to the analog voltage according to the phase-voltage relation shown in Fig. \ref{fig:phase_vs_vr}. Since the conversion from voltage to phase is near-linear, in theory, the analog voltage value can be obtained by simply dividing the phase by a constant. However, random measurement error exists in the CSI phase difference in practice. To deal with this error, we split the whole possible phase range into several discrete segments to achieve digitization of the sensor voltage. The number of segments determines the resolution of this sampling process. A higher segment number means higher throughput, but more errors may be introduced to digitization. We evaluate the digitization resolution of the \SystemName tag in \S\ref{subsec:digitization_resolution}. In this way, \SystemName transfers the sensor readings completely in the analog form without using microprocessors and shifts the sampling part to the WiFi receiver.

Last but not least, according to 802.11n, the ESS CSI is only used as an extra sounding of the wireless channel and is not used to decode the WiFi data. The embedded phase change does not interfere with the decoding of the original payload. Therefore, \SystemName works transparently with WiFi networks with no impact on the WiFi's throughput.

\section{The MAC Layer Design}\label{sec:mac_layer}
As described in the previous section, \SystemName interacts with two WiFi channels (20MHz apart from each other), since frequency shifting is involved. Without confusing the terms, we refer to the channel where the WiFi transmitter operates as the original channel, and the channel that the tag shifts the frequency to as the secondary channel. With regard to the design of MAC (Medium Access Control), \SystemName employs a receiver-initiated process, detailed as follows.

The reader (i.e., the receiver in Fig. \ref{fig:introduction}) switches to the secondary channel when it is ready to receive backscattered sensor data from the tag. The reader first broadcasts a \textit{CTS\_TO\_SELF} to reserve the channel, followed by two consecutive excitation packets with a specific interval. The tag recognizes this pattern of packets with its onboard packet detector. Then the tag wakes up and gets ready for backscatter. From then on, the packets sent by the transmitter in the original channel will excite the tag, making the latter conduct the corresponding operations including phase embedding and frequency shifting.  The reader listening in the secondary channel is able to receive those packets from the tag. The reader can send ACKs back to the transmitter right in the secondary channel. According to the frequency shifting mechanism, the tag will shift the frequency of the ACKs back to the original channel, so that the transmitter can receive them.

As for the transmitter, there is not any modification to its MAC layer. A transmitter can work in a way exactly the same as that in a normal WiFi network.


\section{Implementation}\label{sec:implementation}
\subsection{Backscatter Tag}
We implement the \SystemName tag on a printed circuit board (PCB) using commercial off-the-shelf components, as shown in Fig. \ref{fig:tag_implementation}. The tag contains two RF paths: the packet detector and the backscatter circuit, each connected with a typical 2.4GHz omnidirectional antenna, at 3dBi gain.

The packet detector is implemented using an LT5534\cite{LT5534} envelope detector and a comparator TLV3201\cite{TLV3201}. It identifies the arriving WiFi packet and achieves synchronization so that the tag can locate the HT-ELTF section.
\begin{figure*}[t]
    \centering
    \begin{minipage}{0.49\textwidth}
        \centering
        \subfigure[Throughput]{
            \includegraphics[width=0.47\textwidth]{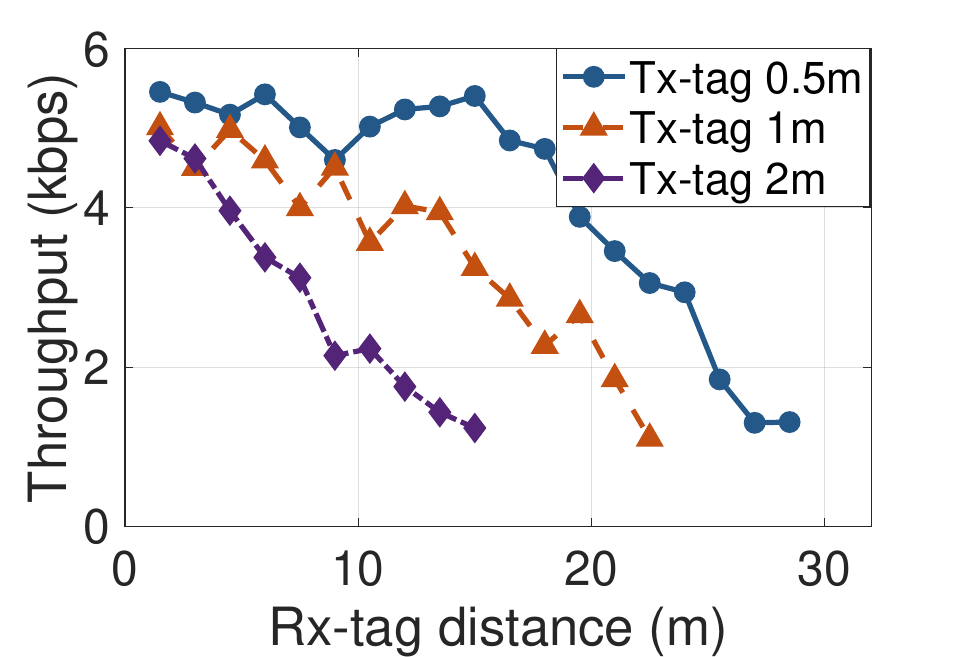}
            \label{exp:throughput_los}
        }
        \subfigure[DER of 10 segments digitization]{
            \includegraphics[width=0.47\textwidth]{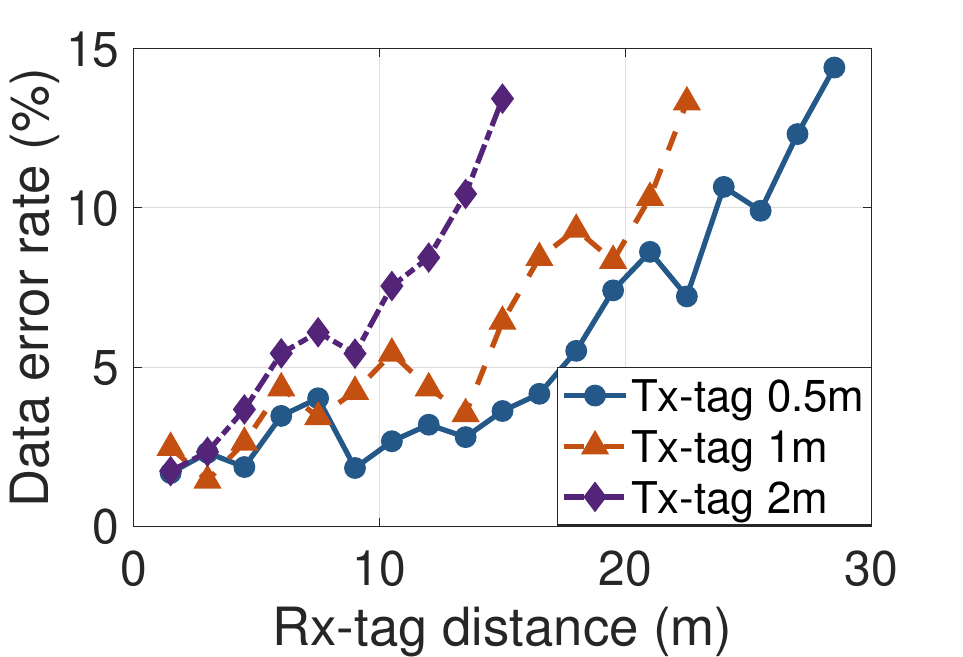}
            \label{exp:ser_los}
        }
        \vspace{-0.2cm}
        \caption{LOS throughput and data error rate.}
        \label{exp:throughput_ser_los}
    \end{minipage}
    \hspace{0.2cm}
    \begin{minipage}{0.49\textwidth}
        \centering
        \subfigure[Throughput]{
            \includegraphics[width=0.47\textwidth]{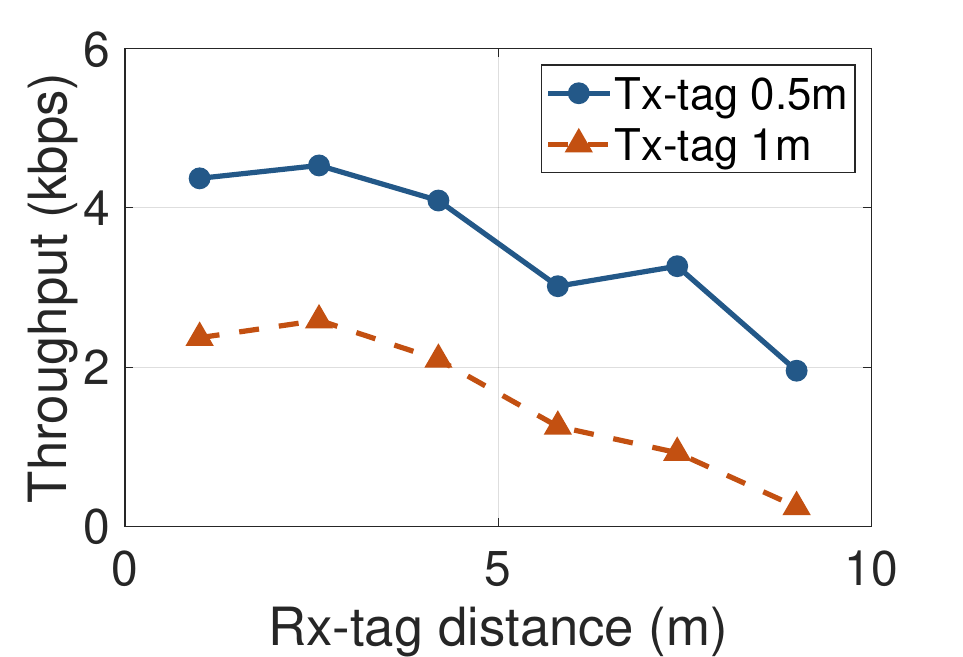}
            \label{exp:throughput_nlos}
        }
        \subfigure[DER]{
            \includegraphics[width=0.47\textwidth]{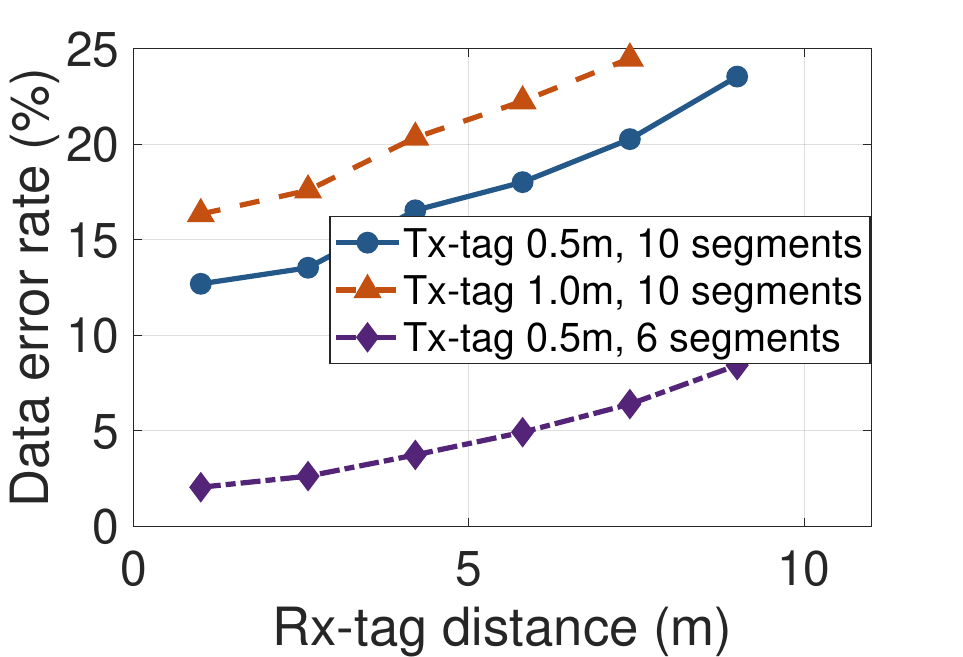}
            \label{exp:ser_nlos}
        }
        \vspace{-0.2cm}
        \caption{NLOS throughput and data error rate.}
        \label{exp:throughput_ser_nlos}
    \end{minipage}
    \vspace{-0.2cm}
\end{figure*}

The backscatter circuit mainly consists of three modules: voltage-phase conversion, control logic, and frequency shifting. We use PathWave ADS to simulate and implement the voltage-phase conversion circuit, using a SMV2201\cite{SMV2201} as the varactor diode and the microstrip line as the transmission line. To switch between the embedding state and the reference state, we use an ADG919\cite{ADG919} RF switch. The control logic of the phase embedding is implemented using a nano-low power ALGN125 FPGA\cite{AGLN125}. \revision{Note that the FPGA included in our prototype only works as a part of the radio, providing its control logic. When delivered to production, the FPGA will be substituted by digital gate circuits, consuming a mere 1 to 2 $\mu$W power.} For frequency shifting, we build a ring oscillator with 3 SN74AUP3G04\cite{SN74AUP3G04} inverters to generate a 20MHz clock. By multiplying the incident signal with this clock signal, the backscattered WiFi signal is moved $\pm$20MHz away on the frequency band. In our implementation, the multiplication is done by toggling an ADG901\cite{ADG901} RF switch at 20MHz.

As an analog backscatter interface, \SystemName tag supports connecting any peripheral analog sensors in a \textbf{plug-and-play} manner. In our implementation, we connect the tag with two types of analog sensors, a light intensity sensor using a photoresistance (GL7549) and a joystick position sensor (EGN-J-O100A). Both sensors output the sensor readings as voltage signals. No MCU is used in such a tag-sensor connection. We show the proof-of-concept application using the light sensor in \S\ref{subsec:poc_app}.

\subsection{WiFi Transceiver}
The WiFi transmitter and the WiFi receiver are two computers equipped with Atheros AR9300 WiFi NICs \cite{9300NIC}. \revision{\SystemName does not modify the WiFi hardware. The transmission of ESS-featured 802.11n packets, CSI acquisition, as well as the aforementioned MAC layer design is enabled after a driver upgrade. Our receiver uses the PicoScenes CSI tool \cite{PicoScenes} to obtain CSI measurements.} It records both the regular CSI and the ESS CSI so that the receiver can extract the embedded sensor readings.


\section{Evaluation}\label{sec:evaluation}
We first show the methodology in \S\ref{subsec:methodology} and the overall performance in \S\ref{subsec:overall_performance}. \S\ref{subsec:power_consumption} presents the power consumption and power benefit of \SystemName. \S\ref{subsec:benchmarks} presents the result of ablation studies. \S\ref{subsec:transparency} evaluates the impact on WiFi carrier transmissions. Finally, we show the proof-of-concept application of \SystemName in \S\ref{subsec:poc_app}.

\subsection{Methodology}\label{subsec:methodology}
To evaluate the transmission capacity and efficiency of \SystemName, we first define its throughput and data error rate. \SystemName's reader splits the possible phase range into several segments. The number of segments determines how many bits can be embedded in a packet. Therefore, the \textbf{throughput} is calculated as the product of the number of received backscatter packets and the number of bits each packet contains. Similarly, a data error occurs when the digitalized sensor voltage falls into the wrong segment. We calculate the \textbf{data error rate (DER)} by measuring the proportion of packets with data errors to the total number of received packets.
\subsection{Overall Performance}\label{subsec:overall_performance}
We first evaluate the throughput and DER of \SystemName in the line-of-sight (LOS) and non-line-of-sight (NLOS) scenarios. We conduct our experiment in an office area, as shown in Fig. \ref{exp:floor_plan}. The transmitter is placed at the end of the corridor and inside a meeting room for line-of-sight and non-line-of-sight scenarios, respectively. It transmits packets at a peak power of 30dBm on WiFi channel 1. The tag  shifts the frequency of the WiFi signal by $\pm$20MHz, so that the backscattered signals are in channel 5. We let the transmitter transmit 2000 packets per second for 10 seconds and receive them at different Tx-tag distances and Rx-tag distances. The experiment in each position is repeated 20 times so that the tag embeds 20 different voltages in the range of 0V to 5V into the WiFi packets. Unless otherwise posted, the following experiments use the same setting.

Upon receiving the packet, the receiver calculates the phase difference in CSIs of all 56 subcarriers. Since the \SystemName tag provides flat phase conversion on the whole frequency band, all 56 subcarriers have the same phases in theory. We take their average to be digitalized. In this experiment, we split the phase range of about 40 degrees into ten segments so that each embedded packet contains about 3.3 bits of information.

The throughput in the LOS scenario is shown in Fig. \ref{exp:throughput_los}. \SystemName achieves a throughput of about 5Kbps when the tag is close to the Tx or the Rx. The throughput decreases when the distance between the tag and Tx/Rx increases due to the energy degradation of the backscattered signal. Moreover, \SystemName achieves a communication range of about 30m, which is comparable to the existing works.

The DER in the LOS scenario is shown in Fig. \ref{exp:ser_los}. We can see that \SystemName achieves a DER of less than 5\% when the tag is close to the Tx or the Rx. The DER increases to about 10\% when the distance increases. Although the phase difference should be fixed in theory, a few degrees of random measurement error exist in the CSI calculation. This random error will increase when the distance increases due to lower signal strength and a more complex multipath environment. To achieve a better DER, we can reduce the number of segments during the digitization, so as to have a higher tolerance of the phase error. We evaluate this method in \S\ref{subsec:digitization_resolution}.

The throughput and DER of \SystemName in the NLOS scenario are shown in Fig. \ref{exp:throughput_ser_nlos}. \SystemName achieves about 4Kbps throughput in the NLOS scenario. For the DER, we find that when using the same segment number as that in the LOS scenario, the DER degrades to around 20\%. Such degradation can be compensated by using smaller segment numbers. When the segment number is 4, the DER will be less than 10\%. In other words, we can trade the throughput for better error tolerance in the NLOS scenario.

The throughput of \SystemName is sufficient to meet the data collection requirement of many IoT sensing applications, but it may be argued that such throughput is not comparable to that of many digital backscatter approaches. It is worth emphasizing that \SystemName achieves such a throughput in a transparent manner, which doesn't hurt the WiFi carrier's throughput, as shown in \S\ref{subsec:transparency}.

\subsection{Power Consumption}\label{subsec:power_consumption}
\subsubsection{Tag Power Consumption}\label{subsubsec:tag_power}
The \SystemName tag's power consumption mainly comes from four parts: RF switches, packet detector, control logic, and the 20MHz clock generation. Here we report the power in an ASIC solution. We use two RF switches in our implementation, one for communication state switching and one for frequency shifting. These switches consume 2$\mu$W in total according to off-the-shelf products \cite{ADG901}. The packet detector can work in a hierarchical mode and  consumes about 7$\mu$W when implemented in 65nm CMOS technology \cite{SyncScatter}. The control logic provides the control signal and consumes 1 to 2$\mu$W power according to existing study \cite{HitchHike}. The clock generation is the major source of power consumption. Similar to the existing WiFi backscatter works \cite{FS-Backscatter}, we use a ring oscillator to generate the 20MHz clock, which consumes 20$\mu$W in ASIC technology. Therefore, the power consumption of a \SystemName tag in the ASIC implementation will be around 30$\mu$W. For a prototype PCB implementation, the power consumption is around 40mW.
\begin{figure}[t]
    \centering
    \includegraphics[width=0.46\textwidth]{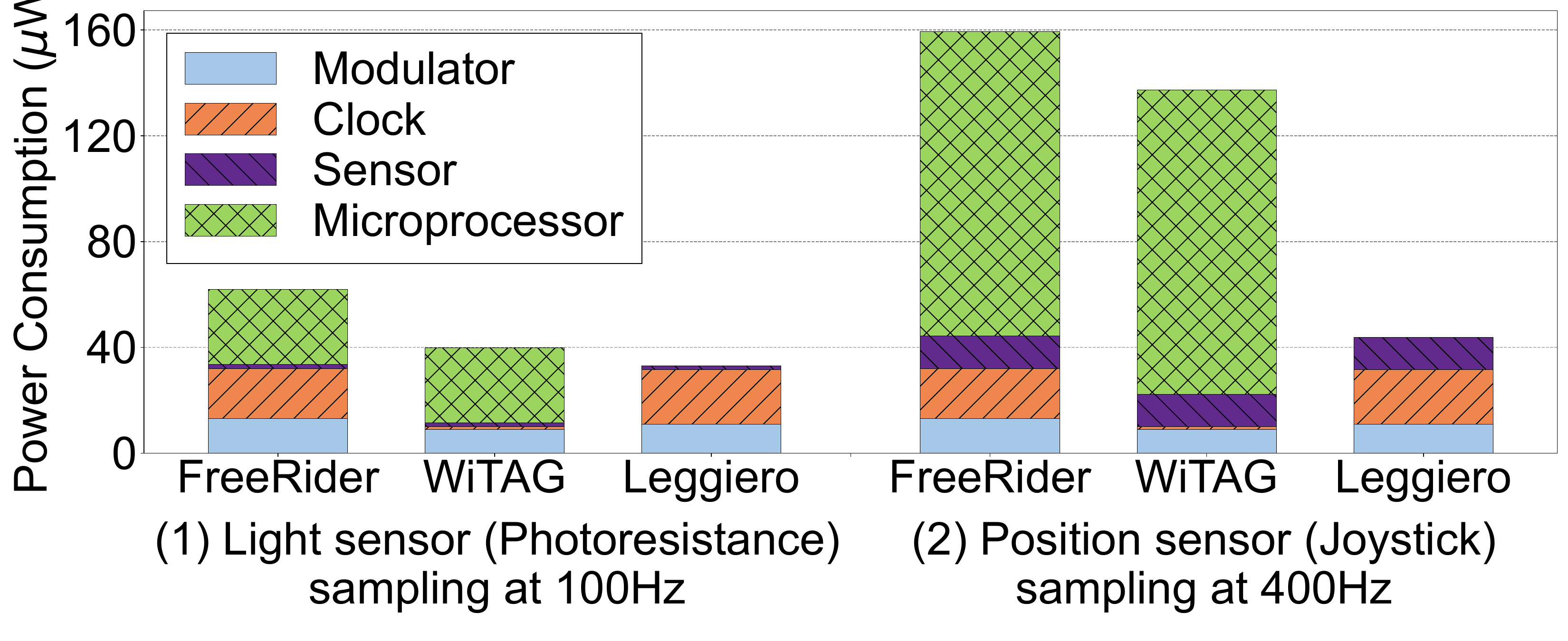}
    \vspace{-0.2cm}
    \caption{Comparison of the end-to-end power consumption breakdown when connecting with sensors.}
    \label{exp:power_breakdown}
    \vspace{-0.4cm}
\end{figure}

Meanwhile, a \SystemName tag can work with existing RF energy harvesting technologies that harvest RF energy. Such an energy harvester can provide more than 30$\mu$W of power \cite{WiFiHarvest}, which is the power consumption of the \SystemName tag. Other energy harvesters such as solar panels can also be considered to power the tag. A small solar panel of 2-3cm$^2$ is sufficient to provide  energy for a \SystemName tag.

\subsubsection{Power Benefit}\label{subsubsec:power_benefit}
\revision{To fully understand the power benefit of \SystemName, we compare it to existing WiFi backscatter systems, specifically FreeRider \cite{FreeRider} and WiTAG \cite{WiTAG}. These two works are representative of state-of-the-art WiFi backscatter systems \cite{VerificationWiFiBackscatter}. We compare the end-to-end power consumption of transferring sensor data, as shown in Fig. \ref{exp:power_breakdown}. We choose power consumption under the same acquisition speed as the metric rather than the energy efficiency of data transmission (measured by bit per joule). This is because we desire to evaluate the entire sensing process of the backscatter tag, including sensor reading acquisition, sensor-radio interfacing, and transmission. Energy efficiency alone fails to take into account the power consumed during acquisition and interfacing.}
\begin{figure}[t]
    \begin{minipage}{0.235\textwidth}\centering
        \includegraphics[width=\textwidth]{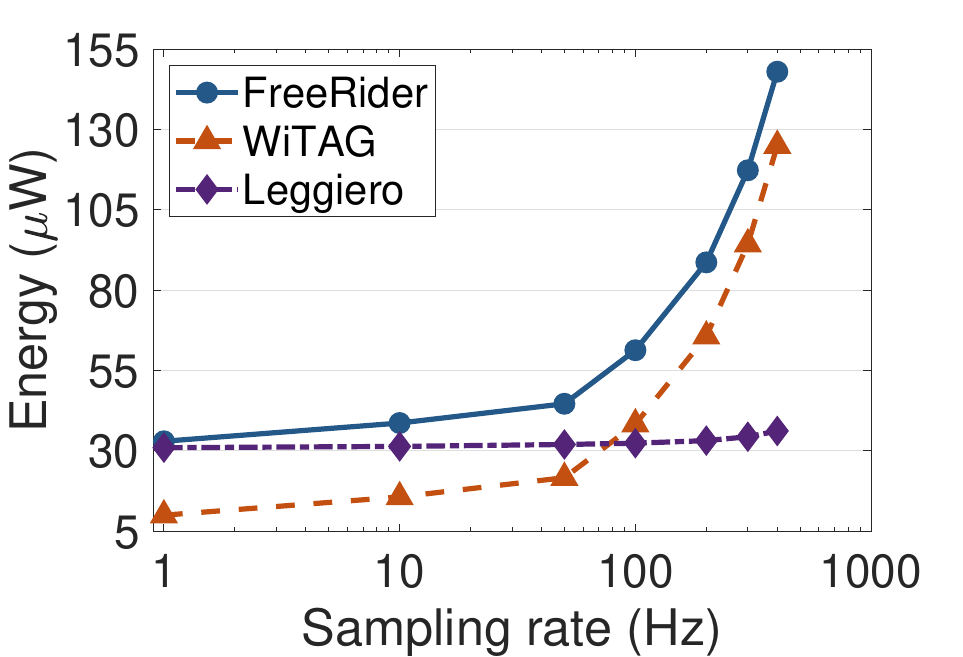}
        \vspace{-0.4cm}
        \caption{Power benefit \\of \SystemName.}
        \label{exp:power_benefit}
    \end{minipage}
    \begin{minipage}{0.235\textwidth}\centering
        \includegraphics[width=\textwidth]{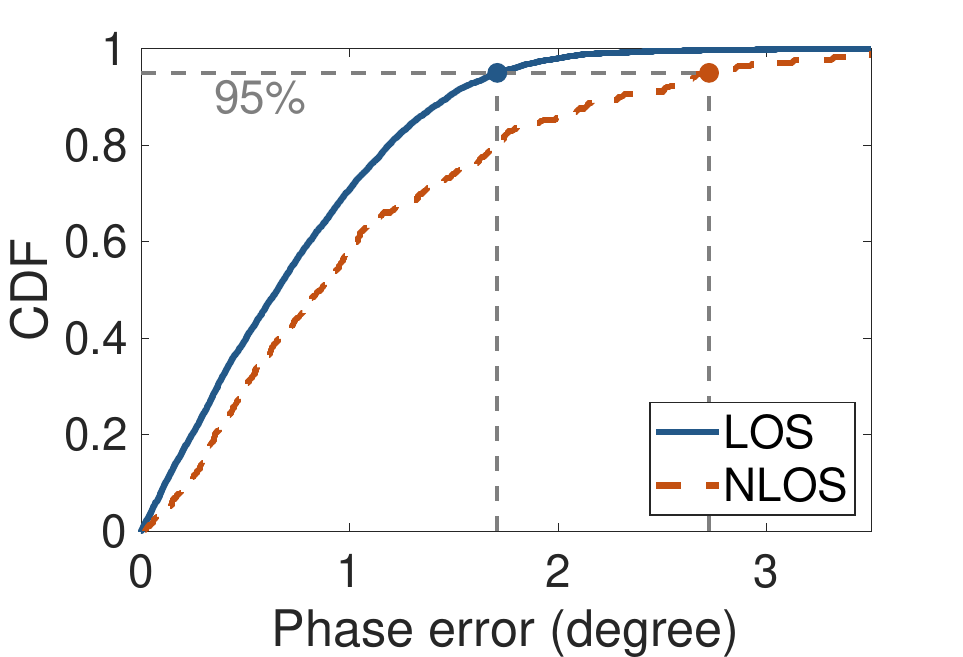}
        \vspace{-0.4cm}
        \caption{Phase error CDF for LOS and NLOS.}
        \label{exp:error_cdf}
    \end{minipage}
    \vspace{-0.4cm}
\end{figure}

\revision{In this experiment, we connect the tags with a light sensor and a position sensor, operating at 100Hz and 400Hz sampling rates, respectively.} The data for WiTAG and FreeRider is acquired and fed to the radio by an external MCU MSP430FR5969 \cite{MSP430FR5969}, which is the common solution for building digital backscatter sensors (e.g., WISP 5.0 platform). This MCU has very low standby and sleep currents (0.4$\mu$A and 0.02$\mu$A) and is put to sleep whenever possible to ensure a fair comparison. \revision{Whereas in \SystemName, the analog sensors can be directly connected. Furthermore, to ensure a fair comparison of the power of the backscatter radio, we use the optimal ASIC implementation result reported in each work \cite{VerificationWiFiBackscatter}}. We can see that \SystemName saves the energy consumption brought by the MCU, which has become the bottleneck for the digital backscatter sensor, especially in the case of a high sampling rate.

To further show the impact of sampling rates, we measure the end-to-end power consumption of the three tags when acquiring and transmitting the light sensor data at different sampling rates. Fig. \ref{exp:power_benefit} shows the result. Excluding the power brought by the peripheral sensor, the \SystemName tag has 4.8$\times$ and 4.0$\times$ lower power consumption at a sampling rate of 400Hz, compared with FreeRider \cite{FreeRider} and WiTAG \cite{WiTAG}, respectively. The two existing approaches require 115$\mu$W power to interface with the sensor, which is often unaffordable for existing RF energy harvesting technologies and solar panels in indoor environments \cite{VerificationWiFiBackscatter}.

\subsection{Ablation Study}\label{subsec:benchmarks}
To better understand the performance of \SystemName, we conduct ablation studies on the phase conversion, analog readings embedding, and readings extraction. \S\ref{subsec:analog_conversion_accuracy} compares the phase conversion between CSI calculation and vector network analyzer (VNA) measurement. \S\ref{subsec:synchronization_error} presents the impact of synchronization errors of the embedding process. \S\ref{subsec:tag_component} shows the design considerations of different reflective circuit components. \S\ref{subsec:digitization_resolution} presents the impact of the digitization resolution on the extraction of sensor readings.

\subsubsection{Analog Conversion Accuracy}\label{subsec:analog_conversion_accuracy}
\begin{figure}[t]
    \begin{minipage}[t]{0.48\textwidth}
        \centering
        \subfigure[Line-of-sight scenario]{
            \includegraphics[width=0.47\textwidth]{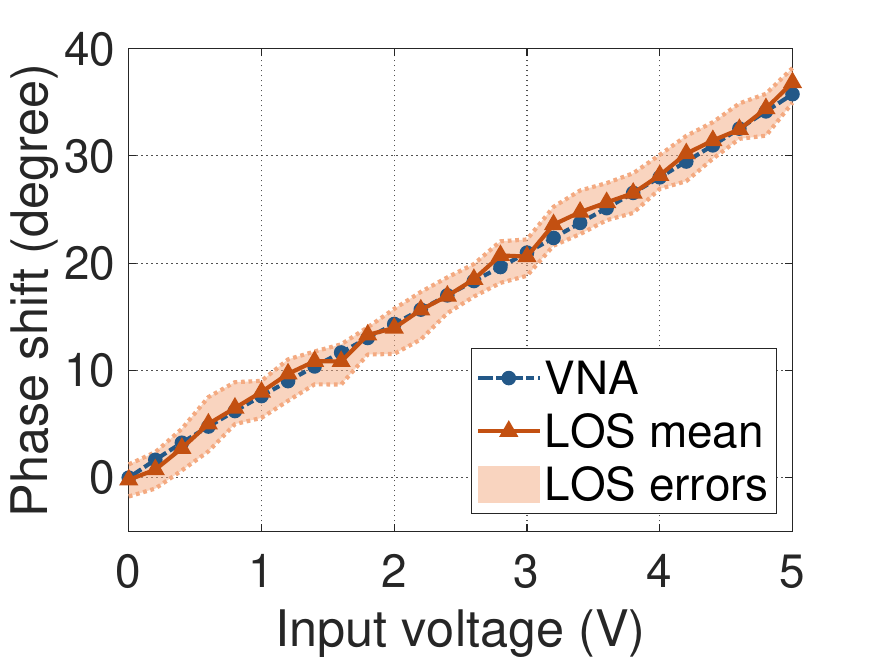}
            \label{exp:conversion_comparison_los}
        }
        \subfigure[Non-line-of-sight scenario]{
            \includegraphics[width=0.47\textwidth]{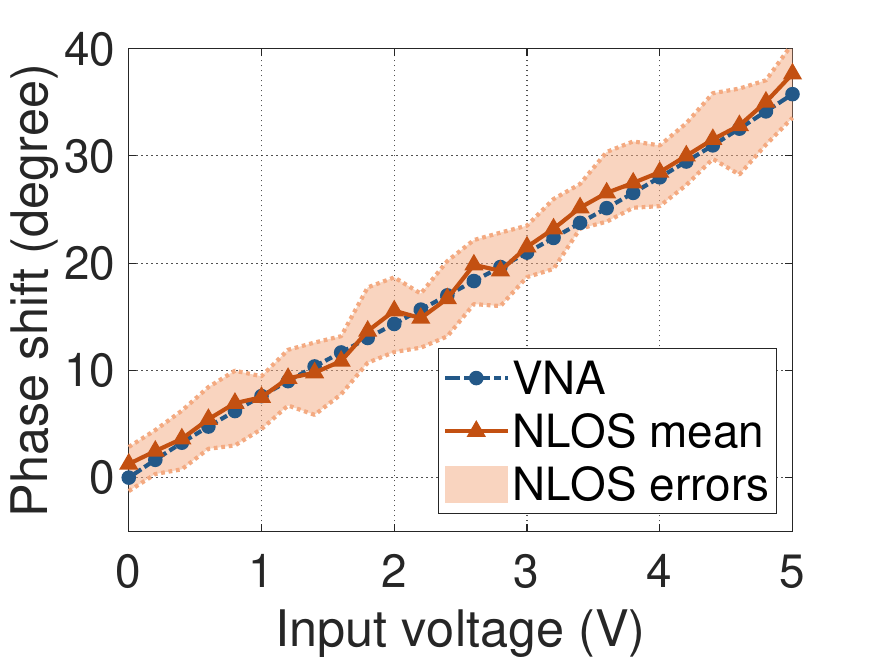}
            \label{exp:conversion_comparison_nlos}
        }
        \vspace{-0.4cm}
        \caption{Phase conversion comparison between VNA and wireless channel.}
        \label{exp:conversion_comparison}
    \end{minipage}
    \vspace{-0.4cm}
\end{figure}

We measure the reflection coefficient of the tag under different input voltages to verify whether it is consistent with our simulation result. We use a Keysight E5071C VNA to measure the $S_{11}$ parameter at an input voltage range of 0V to 5V. The result is shown in the blue line in Fig. \ref{exp:conversion_comparison}. We can see that the \SystemName tag provides a linearly changing phase with respect to the voltage and the phase variation is consistent with the result of the simulation.

Next, we measure the phase-voltage relation of the embedded WiFi CSI. We calculate the phase difference between the ESS CSI and the regular CSI in LOS and NLOS scenarios. The input voltage range is also 0 to 5V with a step interval of 0.2V. Fig. \ref{exp:conversion_comparison_los} and Fig. \ref{exp:conversion_comparison_nlos} show the LOS result and NLOS result, respectively. We can see that the phase embedded in the WiFi CSI changes with the input voltage in a pattern that is the same as the VNA result. However, as we have mentioned, the inevitable measurement error exists in the CSI calculation, and it will result in some phase errors. \revision{Note that these phase errors are solely attributed to the CSI measurement of the channel, and have nothing to do with the tag's phase shifting circuit.} Moreover, the phase error in the NLOS scenario is more than that in the LOS scenario, since the multipath environment is more complex.

To further study the phase error of the embedded WiFi CSI. We show the CDF of phase error in LOS and NLOS scenarios in Fig. \ref{exp:error_cdf}. Although the maximum phase error in the NLOS scenario can be up to 4.5 degrees, more than 95\% of the errors are less than 3 degrees. This means that by carefully selecting the number of segments during the digitization, \SystemName can provide a low DER in the NLOS scenario. \revision{Fig. \ref{exp:error_cdf} also shows that phase errors increase as the multipath becomes more severe. Nevertheless, it is important to note that successful packet decoding is a prerequisite for obtaining the CSI phase. When packets experience extreme multipath conditions, they cannot be successfully received and are naturally excluded from CSI measurement. Therefore, in general, the effect induced by multipath is limited, and the NLOS results in Fig. \ref{exp:error_cdf} exhibit the highest phase errors.}
\subsubsection{Synchronization Error of Embedding}\label{subsec:synchronization_error}
\begin{figure}[t]
    \centering
    \begin{minipage}{0.23\textwidth}\centering
        \includegraphics[width=\textwidth]{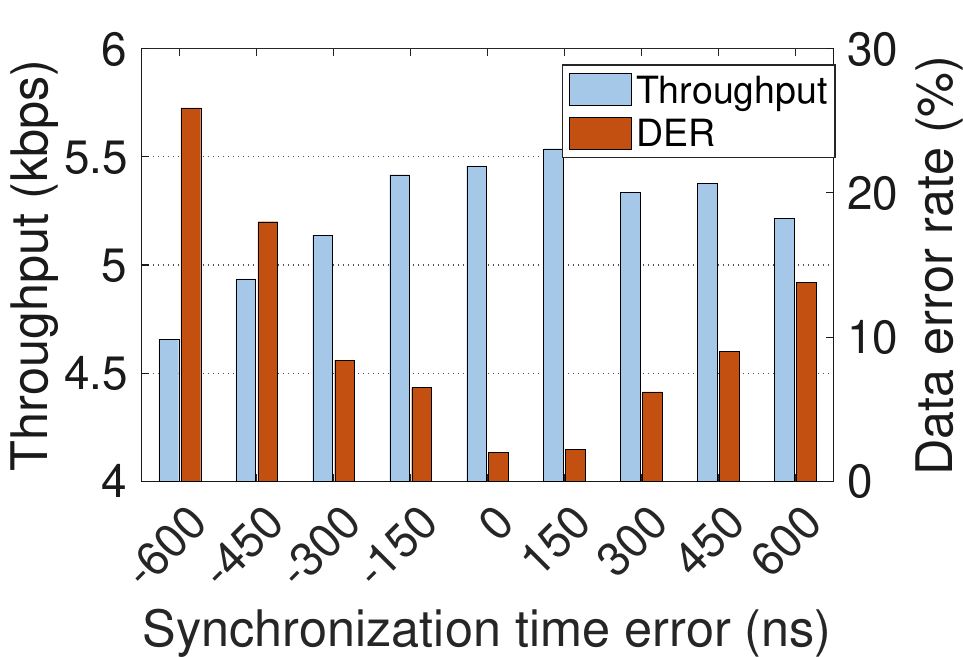}
        \vspace{-0.4cm}
        \caption{Impact of synchronization error.}
        \label{exp:synchronization}
    \end{minipage}
    \hspace{0.1cm}
    \begin{minipage}{0.23\textwidth}\centering
        \includegraphics[width=\textwidth]{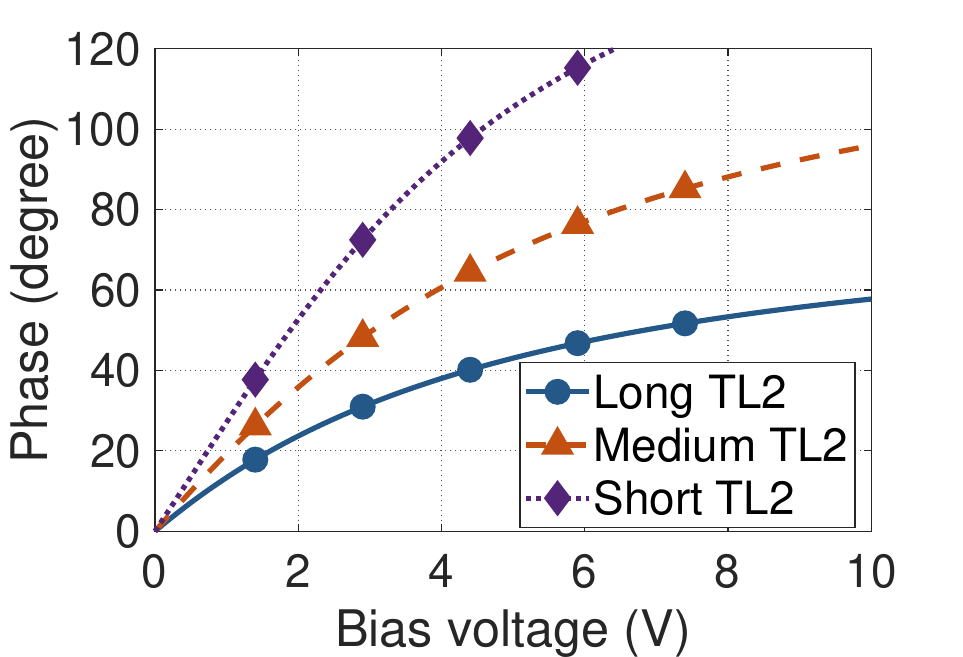}
        \vspace{-0.4cm}
        \caption{Different tag \\component choices.}
        \label{exp:tag_component}
    \end{minipage}
    \vspace{-0.4cm}
\end{figure}

\SystemName precisely embeds the converted phase in the HT-ELTF section of 802.11n packets. When reflecting, the tag needs to synchronize its switching time with this 4$\mu s$ WiFi symbol. We now evaluate the impact of possible synchronization errors. In this experiment, we vary the switching time in a 150$ns$ step to measure the throughput and the DER. The result is shown in Fig. \ref{exp:synchronization}. A negative error means that the state switching of the tag is before the actual arrival time of the HT-ELTF. A positive error means the opposite case. We find that the existence of the synchronization error degrades \SystemName's performance—the greater the error is, the worse the throughput and the DER will be. Interestingly, the degradation brought by the negative and the positive errors is different. A negative error affects the performance more seriously than a positive one with the same absolute value. The reason is that the LTF section contains a 0.8$\mu s$ guard interval (GI) before the actual 3.2$\mu s$ baseband signal. When calculating the CSI, the WiFi receiver only uses the latter 3.2$\mu s$ signal and drops the GI. Therefore, a positive error less than 800 $ns$ still includes the complete 3.2$\mu s$ signal in its embedding state, which results in less degradation. The result shows that \SystemName can tolerate about 300ns synchronization error, which means that the switch requires a 4MHz clock to work properly. Such a clock can be acquired from our ring oscillator with a simple counter.
\subsubsection{Reflective Circuit Component}\label{subsec:tag_component}
In our current implementation, about 40 degrees of maximum phase shift is produced, and the phase changes linearly with the input voltage in the range of 0 to 5V. With different choices of the tag's component, these values may be different. Specifically, the length of the transmission line TL$_2$ in the reflective circuit design, as shown in Fig. \ref{fig:circuitdesign}, can greatly affect the phase-voltage relation. We adjust the length of the TL$_2$ and simulate the reflection coefficient of the circuit. The phase-voltage relation with different lengths is shown in Fig. \ref{exp:tag_component}. We find that by decreasing the length, we obtain a wider range and more fine-grained phase shift. But there are also disadvantages of using a shorter transmission line. The linear phase-voltage variation range will decrease, which means a smaller input voltage range. Moreover, a very short TL$_2$ may lead to a non-flat phase shift on the 2.4GHz frequency band. Given the same input voltage, the phase shift difference between 2.40GHz and 2.50GHz may be up to 10 degrees. Therefore, there is a trade-off in using different transmission lines. Here we leave the tag's component choice to users as they can flexibly select the transmission lines according to the application scenarios.
\subsubsection{Digitization Resolution of Extraction}\label{subsec:digitization_resolution}
When extracting the analog voltage at the tag, the receiver of \SystemName conducts a digitization process. The resolution of this process, namely the number of the split segment, is variable. In this experiment, we show the influence of setting different resolutions. We vary the number of the split segments and measure the throughput and the DER of \SystemName at a fixed distance. The result is shown in Fig. \ref{exp:sampling_resolution}. With a higher segment number, \SystemName achieves higher resolution and therefore higher data throughput. At the same time, the DER goes up. There exists a trade-off in choosing the resolution, and the decision depends on the demand for high throughput or high reliability.
\subsection{Impact on WiFi Carrier's Traffic}\label{subsec:transparency}
\begin{figure}[t]
    \centering
    \begin{minipage}{0.23\textwidth}\centering
        \includegraphics[width=\textwidth]{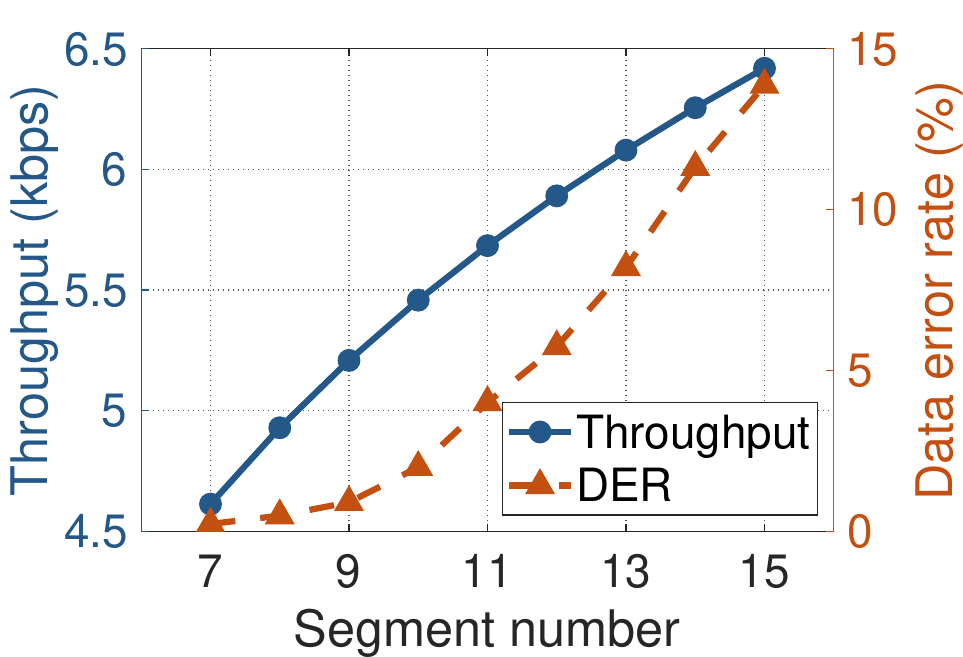}
        \vspace{-0.4cm}
        \caption{Impact of digitization resolution.}
        \label{exp:sampling_resolution}
    \end{minipage}
    \hspace{0.1cm}
    \begin{minipage}{0.23\textwidth}\centering
        \includegraphics[width=\textwidth]{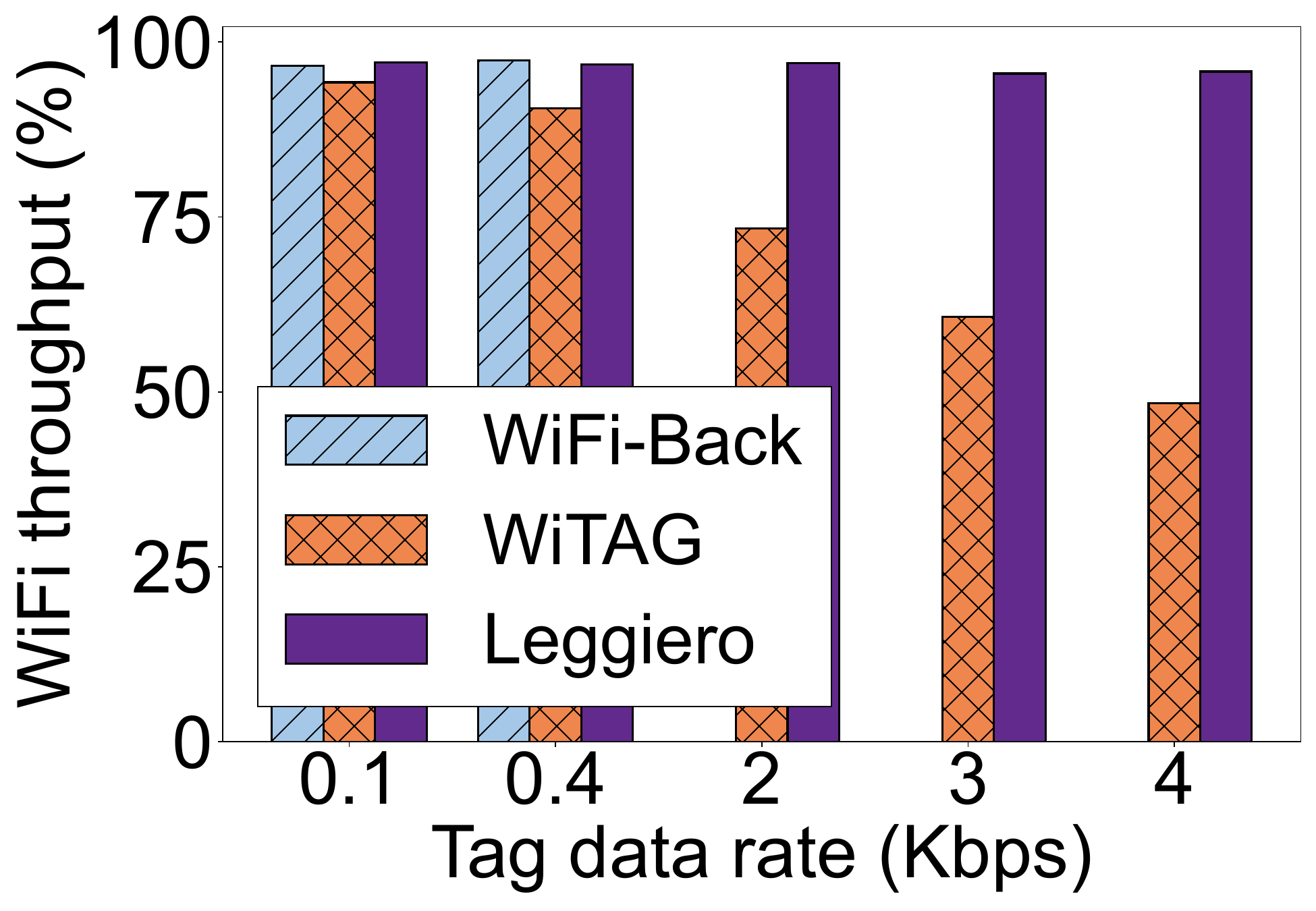}
        \vspace{-0.4cm}
        \caption{Tag's impact on WiFi throughput.}
        \label{exp:transparency}
    \end{minipage}
    \vspace{-0.4cm}
\end{figure}

We compare \SystemName with existing works in terms of the impact on the WiFi carrier's traffic. We consider the single-reader backscatter approaches as the targets to compare. The dual-reader approaches such as HitchHike \cite{HitchHike}, FreeRider \cite{FreeRider}, and MOXScatter \cite{MOXscatter} require two readers to listen to the transmitter simultaneously and thus have low transparency in the coexistence scenario. We make a simulation to measure the reader's maximum WiFi throughput under different tag data rates. We calculate the ratio of the throughput with the backscatter tag and without the tag, as shown in Fig. \ref{exp:transparency}. For \textit{WiFi-Backscatter}\cite{WiFiBackscatter14}, although it has high transparency, it only provides 400bps of data rate. For WiTAG\cite{WiTAG}, since it uses MAC layer OOK modulation, it corrupts about half of the frames to reach its maximum 4Kbps data rate. \SystemName can preserve all the packet payloads and always has high transparency regardless of the tag data rate. \revision{Note that in this experiment, \SystemName's result includes the overhead of using ESS-enabled packets, while regular WiFi packets are used for existing works.}
\subsection{Proof-of-Concept Application}\label{subsec:poc_app}
\begin{figure}[t]
    \begin{minipage}{0.48\textwidth}
        \centering
        \subfigure[Experiment setup]{
            \includegraphics[width=0.31\textwidth]{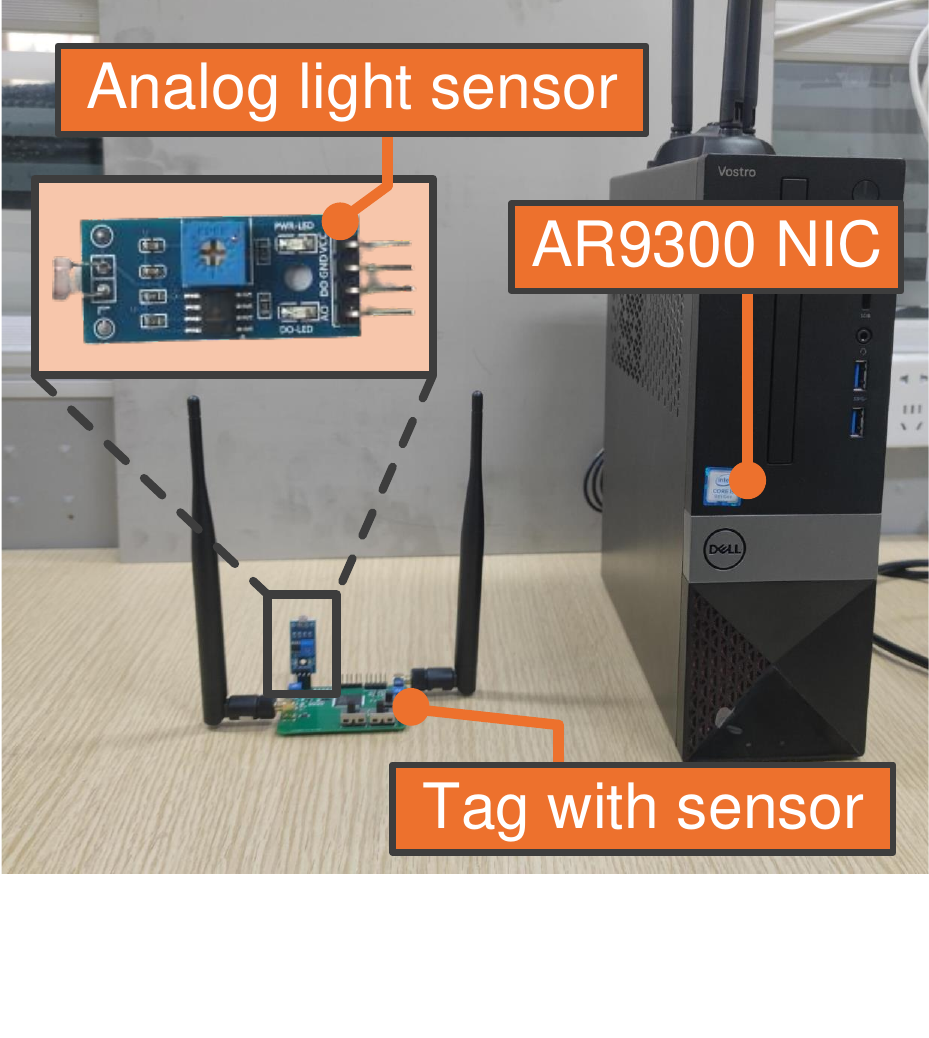}
            \label{exp:poc_setup}
        }
        \subfigure[Light sensor voltage comparison]{
            \includegraphics[width=0.60\textwidth]{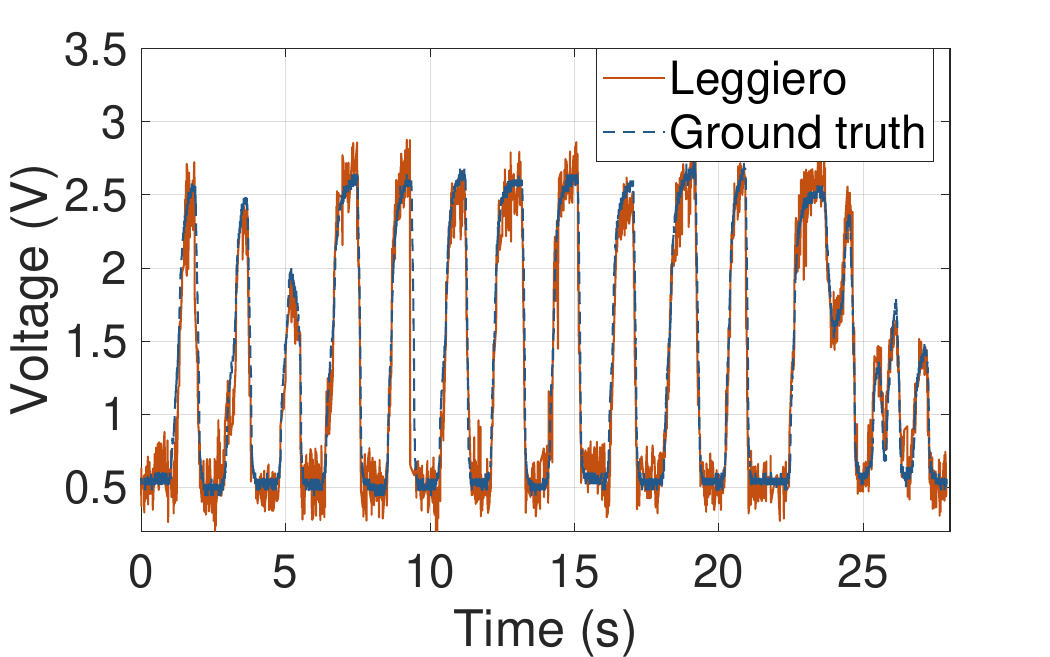}
            \label{exp:poc_app}
        }
        \vspace{-0.4cm}
        \caption{Proof-of-concept application with \SystemName.}
        \label{exp:poc}
    \end{minipage}
    \vspace{-0.4cm}
\end{figure}

We build a real-world application of \SystemName, in which we connect the tag with the low-power light intensity sensor, as shown in Fig. \ref{exp:poc_setup}. We manually modify the ambient light intensity and compare the \SystemName results with the ground truth in real-time, as shown in Fig. \ref{exp:poc_app}. The ground truth voltage is measured directly on the tag by a Tektronix MDO3000 oscilloscope, while the \SystemName voltage is calculated from the CSI phase difference according to the linear relationship shown in Fig. \ref{exp:conversion_comparison_los}. \revision{This experiment is conducted in a noisy office environment to evaluate the robustness of \SystemName, with people walking around and computers and WiFi routers operating near the test site. The tag is placed 1 meter away from the WiFi transmitter and 10 meters away from the receiver.}

We can see deviations in the \SystemName voltages in Fig. \ref{exp:poc_app}, which are caused by the errors in the CSI phase differences shown in Fig. \ref{exp:error_cdf}. Since 95\% of the phase errors are within 2 degrees in the LOS scenario, and 3 degrees in the NLOS scenario. That translates into voltage deviations within 0.25V and 0.375V, respectively. These deviations can be smoothed in the digitization process.

\SystemName directly transmits analog voltage signals over the air. Most types of analog sensors can be easily integrated with our tag without the need for complex interfacing or programming. We believe such a convenient plug-and-play scheme is a promising direction for future IoT systems.

\section{Discussion}\label{sec:discussion}
\textbullet\ \textbf{Post-processing of sensor data.} The analog domain signal conversion of \SystemName is limited to producing the raw sensor signal. In reality, however, post-processing of the sensor data, such as local filtration or aggregation, is often needed in a sensing application. With the current design of \SystemName, such functionalities are offloaded from the local sensor unit to the remote WiFi receiver, which is more powerful in terms of computational capacity. We are also interested in exploring the design of low-power analog processing circuits \cite{AnalogCircuit} to realize such functionalities in our future work.

\revision{\textbullet\ \textbf{Digitization and noise.} We split the entire voltage range into several segments in the digitization process as a way of showing the bit throughput of \SystemName. However, in real-world applications, the sensor produces arbitrary analog voltage values, and no prior segmentation is possible. Instead, the collected sensor readings (i.e., arbitrary voltages) are sampled when the receiver completes reception and can be used directly. This delayed sampling is the key difference \SystemName introduces to sensing applications, as the sampled sensor readings can contain noise from the wireless communication channel. It causes jitters or deviations in the sensor readings as presented in \S\ref{subsec:poc_app}. To mitigate these jitters, a low-pass or Kalman filter can be applied to the sensor readings, similar to the post-processing techniques used in modern IMU sensors \cite{WHEELTEC}. Alternatively, one can decrease the length of the transmission line $TL_2$ according to \S\ref{subsec:tag_component}, to achieve a more fine-grained phase shift and reduce voltage jitters for a fixed CSI phase deviation (e.g., 2 degrees in our LOS experiment).}

\textbullet\ \textbf{Multi-sensor integration.} It is feasible to connect the \SystemName tag with multiple peripheral sensors. The design presented in this paper provides only one port for the sensor connection. It is not difficult to add a peripheral interface that interacts between the analog backscatter radio and the peripheral sensors. A simple solution is to use a multiplexer on the tag and transmit the sensor readings in a round-robin manner.

\textbullet\ \textbf{Multi-tag support.} When there are multiple tags in the network, access to the shared medium can be controlled by the reader. Specifically, the interval between two consecutive wake-up packets in the MAC layer design can be used to indicate the tag's ID. By altering this interval, the reader can switch the channel access from one tag to another.

\textbullet\ \textbf{Generalization of our approach.} Although ESS is enabled only in 802.11n, a legacy LTF field in newer standards (e.g., 802.11ac) also provides duplicate CSI for \SystemName's data embedding. As for MIMO, ESS CSI can work as duplicate channel information for a specific spatial stream to embed the tag's data. Therefore, applying \SystemName to newer WiFi standards and MIMO scenario is feasible.


\section{Related Works}\label{sec:related_works}
IoT sensing systems have experienced significant advances in recent years \cite{IndustrialIoT}, evolving from early wireless sensor networks \cite{Canopy, Duty-CyclingWSN, SmarTiSCH, DistributedWSN} to emerging research areas such as wireless sensing \cite{CSISurvey, AmbiEar, MicNest} and battery-free sensing \cite{OmniTrack, ShopMiner}. Among the battery-free sensing solutions, backscatter technology has emerged as a promising option \cite{BackscatterNetworks, ParallelBackscatter, ParallelBack_Ton, Palantir, Chipnet, PushingRange, Vmscatter, in-Body, RFTransformer}. Traditional backscatter systems such as RFID \cite{RFIDSensing, RED, RFID-Shopping, Versatile_RFID} require specialized readers to communicate with the tag. Ambient backscatter approaches utilize the existing wireless signals such as WiFi \cite{BackFi, PassiveWiFi, SyncScatter, MOXscatter, OFDMAEnabled, RedefiningPassive}, LoRa \cite{PLoRa, Saiyan, LoraBackscatter, Aloba}, LTE \cite{LTEBackscatter}, Bluetooth \cite{InterTechnologyBackscatter, EverySmartPhone, BLEBackscatter}, and even mmWave \cite{mmTag, mmVib} as the carrier signals. \SystemName is related to two categories of backscatter works: WiFi backscatter and analog backscatter.
\begin{table}[t]
    \vspace{0.4cm}
    \resizebox{0.48\textwidth}{!}{
        \begin{tabular}{@{}cccccc@{}}
        \toprule
        \textbf{Technology} & \textbf{\begin{tabular}[c]{@{}c@{}}Through-\\ put\end{tabular}} & \textbf{\begin{tabular}[c]{@{}c@{}}Tag-Rx\\ range\end{tabular}} & \textbf{\begin{tabular}[c]{@{}c@{}}Trans-\\ parency\end{tabular}} & \textbf{\begin{tabular}[c]{@{}c@{}}Power at\\400Hz sampling\end{tabular}} & \textbf{\begin{tabular}[c]{@{}c@{}}Require\\ $\boldsymbol{\mu}$P\end{tabular}} \\  \midrule
        WiFi-Back.\cite{WiFiBackscatter14} & 0.4Kbps & 2m & {\color[HTML]{32CB00} High} & {\color[HTML]{FE0000} 125$\mu$W} & {\color[HTML]{FE0000} Yes}  \\
        HitchHike\cite{HitchHike} & 300Kbps & 50m & {\color[HTML]{FE0000} Low} & {\color[HTML]{FE0000} 147$\mu$W} & {\color[HTML]{FE0000} Yes} \\
        WiTAG\cite{WiTAG} & 4Kbps & 15m  & {\color[HTML]{656565} Med} & {\color[HTML]{FE0000} 125$\mu$W} & {\color[HTML]{FE0000} Yes} \\ \midrule
        \SystemName & 5Kbps & 30m & {\color[HTML]{32CB00} High} & {\color[HTML]{32CB00} 30$\mu$W} & {\color[HTML]{32CB00} No} \\ \bottomrule
        \end{tabular}
    }
    \vspace{0.1cm}
    \caption{\revision{Comparison with existing WiFi backscatter systems.}}
    \label{tb:comparing}
    \vspace{-0.8cm}
\end{table}
\subsection{WiFi Backscatter}
WiFi backscatter takes the ambient WiFi as the carrier signal to transmit data. Some existing WiFi backscatter approaches \cite{FS-Backscatter, HitchHike, FreeRider} propose to separate backscattered traffic from the carrier’s traffic by shifting the backscattered signal's frequency. As a result, they require two receivers simultaneously listening on two channels to decode the backscattered data. Such dual-receiver designs are essentially customized and do not work in an arbitrary WiFi network.

Two existing approaches work with any WiFi transceiver pairs, namely \textit{WiFi-Backscatter}\cite{WiFiBackscatter14} and WiTAG\cite{WiTAG}. \textit{WiFi-Backscatter}\cite{WiFiBackscatter14} modulates tag data on the  WiFi carrier's packets by reflecting or absorbing the WiFi signals. The receiver employs an energy-based detection scheme to decode the tag data, which has gained widespread usage in heterogeneous communication \cite{CTC_Survey}. It is far constrained in terms of throughput (0.4kbps) and communication range (3m). In WiTAG\cite{WiTAG}, the tag corrupts subframes in an aggregated frame from the sender, and the receiver uses the block ACK to transmit data back to the sender. Since block ACK is initially used for ACK from the receiver to the sender, the operation of WiTAG inevitably interferes with normal WiFi transmissions. For example, it assumes the ACKs are always positive in normal communications. In comparison, Leggiero's backscattered traffic is transparent to the carrier WiFi's traffic and achieves higher throughput and longer range.

We summarize the comparison between \SystemName and existing works in Table \ref{tb:comparing}. \revision{Note that the power consumption values listed in the table for a sampling rate of 400Hz do not take into account any peripheral sensor modules, so as to ensure that they are not specific to any particular application.}
All the existing digital backscatter approaches require complicated operations and the help of MCUs to interface with sensors, which induces relatively high power consumption and often overburdens a tag's limited energy budget. In comparison, Leggiero proposes a tailored design for IoT sensors and avoids high power consumption through analog domain signal conversion.
\subsection{Analog Backscatter}
Existing digital backscatter designs do not include external data acquisition and require microprocessors as the interface media. In contrast, analog backscatter fits with sensor signal streaming inherently. Many sensing and streaming applications can be done in an ultra-low power manner. \cite{HybridBackscatter} builds a battery-free analog sensing platform by continuously varying the impedance of the antenna to achieve an amplitude modulation (AM). \cite{BatteryFreeCellphone} demonstrates a battery-free audio communication system also by modulating the amplitude of the signal. There exists a frequency-modulated analog RF backscatter system \cite{RFBandaid}, but the demodulation is still amplitude-based. One problem with these AM systems is that the amplitude of the analog sensor signal is weak and may be easily influenced by noise. Therefore, these approaches often require a high SNR scenario to work properly.

Recent researches begin to seek preferable analog transmission media. Video backscatter \cite{VideoBackscatter} proposes to use the duration of reflecting in one state to achieve pulse width modulation. It builds a video streaming backscatter link to demonstrate the high throughput of such an analog modulation mechanism. The work in \cite{MicrophoneBackscatter} builds a microphone array backscatter that enables concurrent transmission based on this mechanism. Similarly, \cite{SPIbackscatter} presents a direct analog sensor-radio bus using duration-based analog backscatter. Moreover, the RF signal phase is also considered to convey the sensor readings directly \cite{PhaseRFID, WiForce}.

The differences between \SystemName and the existing analog backscatter approaches mainly lie in two aspects. First, \SystemName is compatible and coexists well with commodity networks. It does not require a dedicated exciter or receiver to generate sine tone excitation or decode the backscattered data. Instead, it embeds sensor readings in the extra CSI while preserving the WiFi carrier's traffic. Second, compared with existing RF phase-based designs, \SystemName provides a generic analog interface for various types of sensors at a low cost. The analog signal conversion only costs \$2, without using expensive components like a circulator (more than \$10) or high insertion loss components like a SAW filter (more than 3dB).
\section{Conclusion}\label{sec:conclusion}
This paper presents \SystemName, an analog WiFi backscatter that enables ultra-low power transmission of the IoT sensor data in commodity WiFi networks. \SystemName directly embeds analog sensor readings in the ESS CSI of WiFi packets, avoiding the use of power-hungry microprocessors. At the same time, \SystemName works transparently with WiFi networks. Our evaluations show that \SystemName provides 4.8$\times$ and 4$\times$ power reduction compared to the existing approaches. It achieves a 5Kbps throughput with minimal effect on the WiFi carrier's throughput performance.

\SystemName introduces a novel passive RF computing mechanism that operates on the RF signal’s phase during its propagation in the analog domain. It is low power, low latency, and high efficiency, making it suitable for ubiquitous sensing. Looking ahead, the future work on \SystemName may include incorporating analog data post-processing capabilities and improving the capacity of the passive phase embedding. Moreover, we believe that further research on similar RF computing techniques will facilitate the advancement of low-power IoT technology, and we look forward to exploring these possibilities in future studies.

\begin{acks}
We sincerely thank our anonymous shepherd and all reviewers for their valuable feedback. This work is supported in part by the Joint Funds of the National Natural Science Foundation of China under grant No. U21B2007, and the Guoqiang Institute of Tsinghua University under grant No. 2021GQG1002.
\end{acks}

\balance
\bibliographystyle{ACM-Reference-Format}
\bibliography{refs}


\end{document}